# Big Data Analytics for Wireless and Wired Network Design: A Survey

Mohammed S. Hadi [1,\*], Ahmed Q. Lawey [1], Taisir E. H. El-Gorashi [1] and Jaafar M. H. Elmirghani [1]

[1] *School of Electronic and Electrical Engineering, University of Leeds, United Kingdom*

***Abstract*—Currently, the world is witnessing a mounting avalanche of data due to the increasing number of mobile network subscribers, Internet websites, and online services. This trend is continuing to develop in a quick and diverse manner in the form of big data. Big data analytics can process large amounts of raw data and extract useful, smaller-sized information, which can be used by different parties to make reliable decisions.**

**In this paper, we conduct a survey on the role that big data analytics can play in the design of data communication networks. Integrating the latest advances that employ big data analytics with the networks' control/traffic layers might be the best way to build robust data communication networks with refined performance and intelligent features. First, the survey starts with the introduction of the big data basic concepts, framework, and characteristics. Second, we illustrate the main network design cycle employing big data analytics. This cycle represents the umbrella concept that unifies the surveyed topics. Third, there is a detailed review of the current academic and industrial efforts toward network design using big data analytics. Forth, we identify the challenges confronting the utilization of big data analytics in network design. Finally, we highlight several future research directions. To the best of our knowledge, this is the first survey that addresses the use of big data analytics techniques for the design of a broad range of networks.**

## 1. Introduction

Networks generate traffic in rapid, large, and diverse ways, which leads to an estimate of 2.5 exabytes created per day [1]. There are many contributors to the increasing size of the data. For instance, scientific experiments can generate lots of data, such as CERN's Large Hadron Collider (LHC) that generates over 40 petabyte each year [2]. Social media also has its share, with over 1 billion users, spending an average 2.5 hours daily, liking, tweeting, posting, and sharing their interests on Facebook and Twitter [3]. It is without a doubt that using this activity-generated data can affect many aspects, such as intelligence, e-commerce, biomedical, and data communication network design. However, harnessing the powers of this data is not an easy task. To accommodate the data explosion, data centers are being built with massive storage and processing capabilities, an example of which is the National Security Agency (NSA) Utah data centre that can store up to 1 yottabyte of data [4], and with a processing power that exceeds 100 petaflops [5]. Due to the increased needs to scale-up databases to data volumes that exceeded processing and/or storage capabilities, systems that ran on computer clusters started to emerge. Perhaps the first milestone took place in June 1986 when Teradata [6] used the first parallel database system (hardware and software), with one terabyte storage capacity, in Kmart data warehouse to have all their business data saved and available for relational queries and business analysis [7, 8]. Other examples include the Gamma system of the University of Wisconsin [9] and the GRACE system of the University of Tokyo [10].

In light of the above, the term "Big Data" emerged, and it can be defined as high-volume, high-velocity, and high-variety data that provides substantial opportunities for cost-effective decision-making and enhanced insight through advanced processing which extracts information and knowledge from data [11]. Another way to define big data is by saying it is the amount of data that is beyond traditional technology capabilities to store, manage, and process in an efficient and easy way [12]. Big data is already being employed by digital-born companies like Google and Amazon to help these companies with data-driven decisions [13]. It also helps in the development of smart cities and campuses [14], as well as in other fields like agriculture, healthcare, finance [15], and transportation [16]. Big data has the following characteristics:

1- *Volume*: This is a representation of the data size [17].

2- *Variety*: Generating data from a variety of sources results in a range of data types. These data types can be structured (e.g. e-mails), semi-structured (e.g. log files data from a webpage); and unstructured (e.g. customer feedback), and hybrid data [18].

3- *Velocity*: Is an indication of the speed of the data when being generated, streamed, and aggregated [19]. It can also refer to the speed at which the data has to be analyzed to maintain relevance [17].

Depending on the research area and the problem space, other terms or Vs can be added. For example, is this data of any value? How long can we consider this an accurate and valid data? Since we are conducting a survey, we find it compelling to briefly introduce other Vs as well. Typically, the number of analyzed Vs is 3 to 7 in a single paper (e.g. 6V+C [20]), where C represents *Complexity*, however, different papers analyze different sets of Vs and the union (sum) of all the analyzed Vs among all surveyed papers is 8V and a C, as shown in **Table 1**.

\* Corresponding author.

E-mail addresses: elmsha@leeds.ac.uk (M. Hadi), a.q.lawey@leeds.ac.uk (A. Q. Lawey), T.E.H.Elgorashi@leeds.ac.uk (T. E. H. El-Gorashi), J.M.H.Elmirghani@leeds.ac.uk (J. M. H. Elmirghani)

4- *Value*: Is a measure of data usefulness when it comes to decision making [19], or how much added-value is brought by the collected data to the intended process, activity, or predictive analysis/hypothesis [21].

5- *Veracity*: Refers to the authenticity and trustworthiness of the collected data against unauthorized access and manipulation [21, 22].

6- *Volatility*: An indication of the period in which the data can still be regarded as valid and for how long that data should be kept and stored [23].

7- *Validity*: This might appear similar to veracity; however, the difference is that validity deals with data accuracy and correctness regarding the intended usage. Thus, certain data might be valid for an application but invalid for another.

8- *Variability:* This refers to the inconsistency of the data. This is due to the high number of distributed autonomous data sources [24]. Other researchers refer to the variability as the consistency of the data over time [22].

9- *Complexity*: A measure of the degree of interdependence and inter-connectedness in big data [20]. Such that, a system may witness a (substantial, low, or no) effect due to a very small change(s) that ripples across the system [19]. Also, complexity can be considered in terms of relationship, correlation and connectivity of data. It can further manifest in terms of multiple data linkages, and hierarchies. Complexity and its mentioned attributes can however help better organize big data. It should be noted that complexity was included among the big data attributes (Vs) in [20] where big data was characterized as having 6V + complexity. This is how we will arrange it in **Table 1**.

The process of extracting hidden, valuable patterns, and useful information from big data is called *big data analytics* [44]. This is done through applying advanced analytics techniques on large data sets [28]. Before commencing the analytics process, data sets may comprise certain consistency and redundancy problems affecting their quality. These problems arise due to the diverse sources from which the data originated. *Data pre-processing* techniques are used to address these problems. The techniques include integration, cleansing (or cleaning), and redundancy elimination, and they were discussed by the authors in [39].

Big data analytics can be carried out using a number of frameworks (shown below) that usually require an upgradeable cluster dedicated solely for that purpose [17]. Even if the cluster can be formed using a number of commodity servers [45], however, this still forms an impediment for limited-budget users who want to analyze their data. The solution is presented through the democratization of computing. This made it possible for any-sized company and business owners to analyze their data using cloud computing platforms for big data analytics. Consequently, the use of big data analytics is not limited to enterprise-level companies. Furthermore, business owners do not have to heavily invest in an expensive hardware dedicated to analyzing their data [1]. Amazon is one of the companies that provide 'cloud-computed' big data analytics for its customers. The service is called Amazon EMR (Elastic MapReduce), and it enables users to process their data in the cloud with a considerably lower cost in a pay-as-you-use fashion. The user is able to shrink or expand the size of the computing clusters to control the data volume handled and response time [1, 46].

Dealing with big amounts of data is not an easy task, especially if there is a certain goal in mind since data arrives in a fast manner, it is vital to provide fast collection, sorting, and processing speeds. Apache Hadoop was created by Doug Cutting [47] for this purpose. It was later adopted, developed, and released by Yahoo [48]. Apache Hadoop can be defined as a top-level, java-written, open source framework. It utilizes clusters of commodity hardware [49].

Hadoop V1.x (shown in **Fig.1**) consists of two parts: the Hadoop Distributed File System (HDFS) that consists of a storage part, and a data processing and management (MapReduce) part. The master node has two processes, *a Job Tracker* that manages the processing tasks and a *Name Node* that manages the storage tasks [50].

When a Job Tracker takes job requests, it splits the accepted job into tasks and pushes them to the *Task Trackers* located in the slave nodes [51]. The *Name Node* resembles the master part, while the *Data Nodes* represent the slave part [12]. There is more explanation in the HDFS part below.

Many projects were developed in a quest to either complement or replace the above parts, and not all projects are hosted by the Apache Software Foundation, which is the reason for the emergence of the term *Hadoop ecosystem* [47].

Table 1: Various big data dimensions.

| No. of Vs | References | Dimensions (Characteristics) | | | | | | | | |
|---|---|---|---|---|---|---|---|---|---|---|
| | | Volume | Velocity | Variety | Veracity | Value | Variability | Volatility | Validity | Complexity |
| 3Vs | [25-31] | √ | √ | √ | | | | | | |
| 4Vs | [4, 32-34] | √ | √ | √ | √ | | | | | |
| | [35-39] | √ | √ | √ | | √ | | | | |
| 5Vs | [3, 11, 21, 40, 41] | √ | √ | √ | √ | √ | | | | |
| 6Vs | [20, 22, 24, 42] | √ | √ | √ | √ | √ | √ | | | |
| 7Vs | [23, 43] | √ | √ | √ | √ | √ | | √ | √ | |

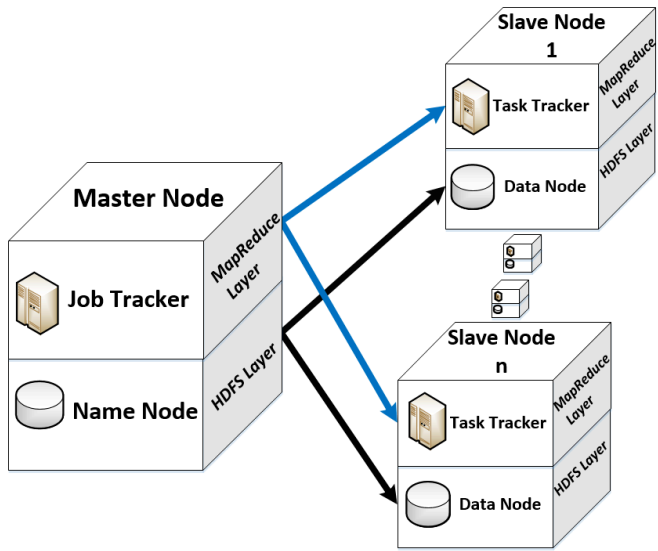

**Fig. 1.** Hadoop V1.x architecture.

Hadoop V2.x is viewed as a three-layered model. These layers are classified as storage, processing, and management, as shown in **Fig. 2.** The current Hadoop project has four components (modules), which are MapReduce, the HDFS, Yet Another Resource Negotiator (YARN), and Common utilities [17].

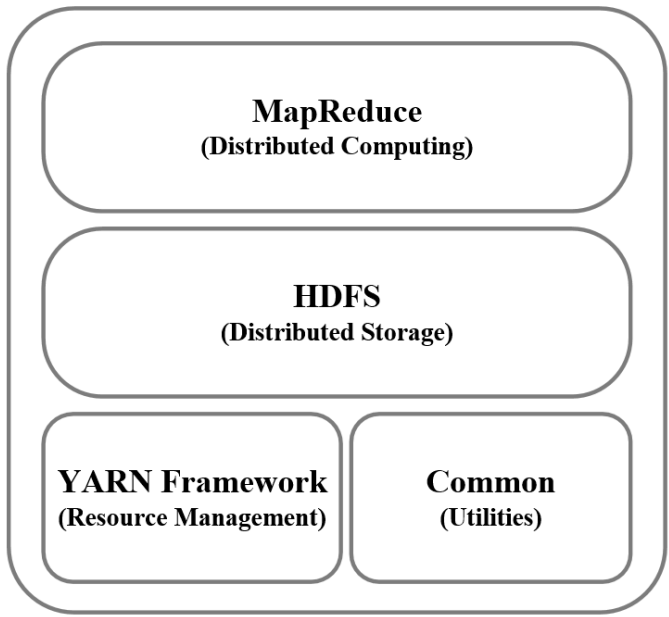

**Fig. 2.** Hadoop V2.x architecture.

1- *MapReduce*: As a programming model, MapReduce is used as a data processing engine and for cluster resource management. With the emergence of Hadoop v2.0, the resource management task became YARN's responsibility [17]. WordCount is an example illustrating how MapReduce works. As the name implies, it calculates the number of times a specific word is repeated within a document. Tuples $\langle w, 1 \rangle$ are produced by the map function, where $w$ and 1 represents the word and the times it appeared in the document respectively. The reduce function groups the tuples that share the same word and sums their occurrences to reach the concluding result [61].

2- *HDFS*: HDFS represents the storage file-system component in the Hadoop ecosystem. Its main feature is to store huge amounts of data over multiple nodes and stream those data sets to user applications at high bandwidth. Large files are split into smaller 128 MB blocks, with three copies of each block of data to achieve fault tolerance in the case of disk failure [17, 52, 53].

3- *YARN*: YARN was introduced in Hadoop version 2.0, and it simply took over the tasks of cluster resource management from MapReduce and separated it from the programming model, thus making a more generalized Hadoop capable of selecting programming models, like Spark [54], Storm [55], and Dryad [56, 57].

4- *Common utilities*: To operate Hadoop's sub-projects or modules, a set of common utilities or components are needed. Shared libraries support operations like error detection, Java implementation for compression codes, and I/O utilities [17, 58].

Over the last few years, researchers in telecommunication networks started to consider big data analytics in their design toolbox. Characterized by hundreds of tunable parameters, wireless network design informed by big data analytics received most of the attention, however, other types of networks received increasing attention as well.

The vast amount of data that can be collected from the networks, along with the distributed modern high-performance computing platforms, can lead to new cost-effective design space (e.g. reducing total cost of ownership by employing dynamic Virtual Network Topology adaptation) when compared to classical approaches (i.e. static Virtual Network Topologies) [59]. This new paradigm is promising to convert networks from being sightless tubes for data into insightful context-aware networks. Our contributions in this paper are as follows:

1- We show in this paper the role big data analytics can play in wireless and wired network design.

2- The above role is corroborated through the illustration of case studies in Section 2.

3- The significance of this paper lies in helping academic researchers save much effort by understanding the state-of-the-art and identifying the opportunities, as well as the challenges facing the use of big data analytics in network design.

4- In addition to academic approaches, we surveyed network equipment manufacturing companies highlighting network solutions based on big data analytics. We also identified the common areas of interest among these solutions, and thus this survey can benefit both academic and industrial-oriented readers.

5- This paper provided insights on potential research directions as illustrated in Section 8.

This paper is organized as follows: Section **2** presents several case studies uses big data analytics in wireless and wired networks. Sections **3-6** illustrate the research conducted in the direction of employing big data analytics in the fields of cellular, SDN & intra-data center, optical networks, and network security, respectively. Section **7** summarizes some of the main big data-based network solutions offered by industry. Section **8** discusses the network design cycle based on big data analytics and highlights the challenges encountered in big data-powered network design. In Section **9** we propose open directions for future research. Finally, the paper ends with conclusions in Section **10**.

## 2. Case studies of the use of big data analytics for wireless and wired networks.

### 2.1 *Detection of sleeping cells in 5G SON*

A wireless cell may cease to provide service with no alarm triggered at the Operation and Maintenance Center (OMC) side. Such cells are referred to as sleeping cells in self organizing networks (SON). The authors in [60] tackled this problem and presented a case study on the identification of the Sleeping Cells (SC). The simulation scenario comprised of 27 macro sites each with three sectors. The user equipment (UE) is configured to send radio measurement and cell identification data of the serving and neighboring cells to the base station, in addition to event-based measurements. The above-mentioned measurements are sent periodically (i.e. every 240 ms). The simulation considered two scenarios; reference (a normally-operating network) and SC. The latter was simulated by dropping the antenna gain from 15 dBi (reference scenario) to -50 dBi (SC scenario). Measurements reported from UEs are then collected from each scenario and stored in a database. The reference scenario provided measurements used by an anomaly detection model that is based on k-nearest-neighbor algorithm to provide a network model with normal behavior. Multidimensional Scaling (MDS) is used to produce a minimalistic Key Performance Index (KPI) representation. Thus the interrelationship between Performance Indexes (PIs) is reflected and an embedded space is constructed. Consequently, similar measurements (i.e. normal network behavior) lie within close distances while dissimilar measurements (i.e. anomalous network behavior) are far-scattered and hence easily identified. The model attained 94 percent detection accuracy with 7 minutes training time.

### 2.2 *A proposed architecture for fully automated MNO reporting system.*

Mobile Network Operators (MNOs) collect vast amounts of data from a number of sources as it can offer actionable plans in terms of service optimization. Visibility and availability of information is vital for MNOs due to its role in decision making. Employing a reporting system is pivotal in the cycle of transforming data to information, knowledge, and lastly to actionable plans. The authors in [61] presented a case study aimed at illustrating the potential role of big data analytics in the development a fully automated reporting system. A Moroccan MNO is to benefit from the alternative architecture. The authors highlighted the shortcomings of the existing automatic reporting system that uses traditional technologies. Moreover, they inferred that using big data analytics can provide the opportunity to overcome those shortcomings.

The authors chose the Apache Flink [61] in their proposed architecture to serve as their big data analytics framework. Several reasons contributed towards this choice, including the Apache Flink's ability to process data in both stream and batch modes, ease of deployment, and fast execution when compared to other frameworks such as Spark. Furthermore, the apache Flink can be integrated with other projects like HDFS for data storage purposes. Moreover, Apache Flink is scalable which makes it an optimal choice for this system.

### 2.3 *Network anomaly detection using NetFlow data*

Big data analytics can support the efforts in the subject of network anomaly and intrusion detection. Towards that end, the authors in [62] proposed an unsupervised network anomaly detection method powered by Apache Spark cluster in Azure HDInsight. The proposed solution uses a network protocol called NetFlow that collects traffic information that can be utilized for the detection of network anomalies. The procedure starts by dividing the NetFlows data embedded in the raw data stream into 1 minute intervals. NetFlows are then aggregated according to the source IP, and data standardization is carried out. Afterwards, a k-means algorithm is employed to cluster (according to normal or abnormal traffic behavior) the aggregated NetFlows. The following step is to calculate the Euclidean distance between the cluster center and its elements. The procedure concludes by evaluating the success criteria. The authors considered a dataset containing 4.75 hours of records captured from CTU University to analyze botnet traffic. The proposed approach attained 96% accuracy and the results were visualized in 3D after employing Principal Component Analysis (PCA) to attain dimension reduction.

## 3. Role of big data analytics in cellular network design

In this section, we review the research done on the use of big data analytics for the design of cellular networks. Compared to other network design topics, we observed that the wireless field has received the most attention, as measured by its share of research papers. These papers can be classified according to the application or area under investigation. Consequently, we have classified those papers into the following:

1- Counter-failure-related: This includes fault tolerance (i.e. detection and correction), prediction, and prevention techniques that use big data analytics in cellular networks.

2- Network monitoring: This illustrates how big data analytics can be beneficial as a large-scale tool for data traffic monitoring in cellular networks.

3- Cache-related: Investigates how big data analytics can be used for content delivery, cache node placement and distribution, location-specific content caching, and proactive caching.

4- Network optimization: Big data analytics can be involved in several topics including predictive wireless resource allocation, interference avoidance, optimizing the network in light of Quality of Experience (QoE), and flexible network planning in light of consumption prediction.

It should be noted that **Table 2** provides further detailed classification, with the chance to compare the role played by big data analytics across different network types and applications.

### 3.1 Failure prediction, detection, recovery, and prevention

#### 3.1.1 Inter-technology failed handover analysis using big data

One of the most frustrating encounters happens when a mobile subscriber gets surprised by a sudden call drop. Many of these incidents occur when the user is at the edge of a coverage area and moving towards another, technologically-different area, e.g., moving from a 3G Base Station (BS) to a 2G BS. The common solutions to address such shortcomings are by either conducting drive tests or performing network simulation. However, another solution that leverages the power of big data was proposed by the authors in [63]. The proposed solution uses big data analytics (Hadoop platform) to analyze the Base Station System Application Part (BSSAP) messages exchanged between the Base Station Subsystem (BSS) and Mobile Switching Center (MSC) nodes. Location updates (only those involved in the inter-technology handover) are identified and the geographic locations where the 3G-service disconnections occur are identified by relying on the provided target Cell ID.

The results of the above method were then compared with a drive test (which is an expensive and time-consuming approach) results, where coherence between the two results was demonstrated. Another comparison was conducted with the Key Performance Index (KPI)-based approach and the results were in favor of the proposed approach.

#### 3.1.2 Signaling data-based intelligent LTE network optimization

By utilizing the combination of all around signaling and user and wireless environment data, combined with Self-Organized Network technologies (SON), full-scale automatic network optimization could be realized.

The authors of [27] developed an intelligent cellular network optimization platform based on signaling data. This system involves three main stages:

1- *Defining network performance indicators through the extraction of XDR keywords*: The External Data Representation (XDR) contains the key information of the signaling (e.g., the causes of the process failures and signaling types). The status of a complete signaling process can also be identified by the XDR (e.g., the success or failure of signaling establishment and release). A number of performance indicators are defined by relying on this information. Querying these indicators is possible from multiple dimensions and levels (e.g., user, cell, and grid level).

2- *Problem discovery:* Service establishment rate, the handover success rate, and drop rate are among the network signaling-plane statuses that can be reflected by the XDR-based network performance indicators. Network equipment with unsatisfactory performance indicators can be further analyzed, and this can be done by conducting a further excavation of the corresponding indicators' original signaling.

3- *Providing best practice solutions*: Identified and solved problems can provide an optimization experience. As a consequence, a variety of network problems can be verified. For example, when a cell has a low handover success rate, according to the definition of the associated indicators, the reason is suggested to be the low success rate of the handover preparation. The solution would be to adjust the overlapping coverage areas formed between the source and the target cells and the parameters (e.g., the decision threshold offset and the handover initiation).

A recommended solution can be provided when a deteriorating indicator surfaces, and this is simply done by clicking the index query that caused the deterioration.

#### 3.1.3 Anomaly detection in cellular networks

When a certain problem occurs in the cellular network, the user would usually be the first who feels the service disruption and suffers the impact. An abnormal and disrupted service may be identified by examining the Call Detail Record (CDR) of the users in a specific area. CDR files are generated upon making a call, and include, among other information, the caller and called numbers, the call duration, the caller location, and the cell ID where the call was initiated or received.

A CDR based Anomaly Detection Method (CADM) was proposed by the authors in [64]. CADM was used to detect the anomalous behavior of user movements in a cellular network. This was done, first, with the CDR data being collected from the network nodes and stored in a mediation department. Then, the second phase starts by distributing the collected CDRs to the relevant departments (e.g., data warehouse, billing, and charging departments). After that, the Hadoop platform is used to detect the anomalies. The discovered anomalies are then fed-back to the mediation department for adequate actions.

The use of big data analytics was essential in this case. Large datasets that require distributed processing across computer clusters were processed by the Hadoop Platform. The result was an improved system that is able to detect location based anomalies and improve the cellular system's performance.

#### 3.1.4 Self-healing in cellular networks

The idea to develop a system that is capable of monitoring itself, detecting the faults, performing diagnoses, issuing a compensation procedure, and conducting a recovery is very appealing. However, the self-healing process has another factor to keep in mind, which is time. The process should be carried out within a reasonable amount of time so it would not degrade the quality of the delivered services.

Three use cases were presented by the authors in [65] for a self-healing process in cellular networks:

1- *Data Reduction:* The Operation and Maintenance (O&M) database can be used for troubleshooting purposes. However, the database size is relatively large as it contains the data related to both normal and degraded intervals, which makes it difficult to process. Separating the intervals to just keep the degraded intervals will help in reducing that size. The authors proposed parallelizing this process independently by analyzing each BS separately.

They chose the degraded interval detection algorithm of [66] (a degraded interval is the time where the BS behavior is degraded), and these intervals were detected by comparing the BS's KPIs to a certain threshold. This algorithm was parallelized by implementing it as a *map* function, a field is added to identify each BS, and all the fields are added by a *reduce* function.

2- *Detecting Sleeping Cells:* Cell outage or sleeping cells is a common problem in mobile networks. Users are directed to neighboring cells instead of the nearest and optimal cell. According to the algorithm described in [67], sleeping cells can be detected through the utilization of neighboring BS measurements hence calculating the impact of the sleeping cell outage. The detection process relies on the Resource Output Period (ROP), where each BS produces Configuration Management (CM), Fault Management (FM), and Performance Management (PM) data every 15 minutes. For each BS, incoming handovers from neighboring BSs are aggregated for the current and previous ROP. If the number of handovers suddenly dropped to zero, and a malfunction is indicated by the cell's Performance Indicators (PIs), the cell is regarded as a sleeping cell.

The authors in [65] proposed the use of the above-mentioned algorithm under the big data principle. They proposed to divide the terrain into partitions that are the maximum distance between neighbors, where each BS within the partitioned area is sequentially tested by an instance of the algorithm, and this is done by examining the data of its neighbors.

This approach was compared to other methods (e.g., lack of KPIs and availability of KPIs), and most of the simulated outages were detected (5.9% false negatives and 0% false positives). While a lack of KPIs and availability of KPIs methodologies showed a high percentage of false negatives.

3- KPI Correlation-Based Diagnosis: The authors in [65] used a method that utilizes most correlated KPIs to identify the problem cause. To simplify the analysis task, the algorithm considers the PIs of both the affected BS and the neighboring sectors.

MapReduce was used to implement this algorithm in a parallelized manner, the correlation process and the creation of a PIs list arranged by correlation were implemented as map and reduce functions, respectively.

### 3.1.5 Cell site equipment failure prediction

A sudden outage of services might have serious consequences, and this is why keeping communication equipment, like cell sites, in a good working state is of high importance. The challenge identified by the authors in [68] is to analyze the user's bandwidth on the cell level. Equipment(s) failure and infrastructure faults can be predicted by analyzing the bandwidth trends in a particular cell.

Due to the size and diversity of the collected data, it is essential to use big data analytics to process it. Thus, the customers' received bandwidth can be acquired over a particular time period (i.e., month or year, etc.). Next the data from diverse data sources are integrated and then analyzed to know the bandwidth trends.

### 3.2 Network monitoring

#### 3.2.1 Large-scale cellular network traffic monitoring and analysis

Large cellular networks have relatively high data rate links and high requirements to meet. Usually these networks use a high-performance and large capacity server to perform traffic monitoring and analysis.

However, with the continuous expansion in data rates, data volumes, and the requirements for detailed analysis, this approach seems to have a limited scalability. Hence, the authors of [69] proposed a system to undertake that task, utilizing the Hadoop MapReduce, HDFS, and HBase (a distributed storage system that manages the storage of structured data and stores them in a key/value pair) as an advanced distributed computing platform. They exploited its capability of dealing with large data volumes while operating on commodity hardware. The proposed system was deployed in the core side of a commercial cellular network, and it was capable of handling 4.2 TB of data per day supplied through 123 Gbps links with low cost and high performance.

#### 3.2.2 Mobile internet big data operator

China Unicom, China's Largest WCDMA 3G mobile operator with 250 million subscribers in 2012, introduced an industry ecosystem. The researchers in [70] highlighted this as a telecom operator-centric ecosystem that is based on a big data platform.

The above-mentioned big data platform is developed for retrieving and analyzing data generated by mobile Internet users. In an aim to optimize the storage, enhance the performance, and accelerate the database transactions, the authors proposed a platform that uses HDFS for distributed storage. The cluster had 188 nodes used to store data, perform statistical data analyses, and as management nodes. The approximate storage space was 1.9 PB. HBase has the role of the distributed database, with a writing rate that can reach 145k records per second; HBase stores the structured data located on the HDFS.

Compared with the Oracle database, it is noted that the system achieved a four times lower insertion rate. The query rate was compared to an Oracle database as well, and the HBase showed a better performance when taking into consideration the impact imposed by the records' size.

### 3.3 Cache and content delivery

#### 3.3.1 Optimized bandwidth allocation for content delivery

Mobile networks, usually, have a large number of users, and with the increase in Internet-based applications, it has become essential to allocate the required bandwidth that meets the user expectations, as well as to ensure a competitive level of service quality. Cellular networks can provide Internet connectivity to their users at any time; however, video (especially high quality) contents are still slow and relatively expensive. From the base station's point of view, the impact of forwarding the same video content to several users on the same base station is massive. The LTE system addressed this through multicast techniques. However, multicast is still regarded as a big challenge in cellular networks. To overcome the above problem, the authors of [26] proposed a solution that can dynamically allocate bandwidth. The idea is based on sharing the base station's wireless channel by a user cluster that wishes to download the contents. Thus, saving the base station resources, as well as providing a better data rate for the clustered users, and providing an opportunity for the users who did not join the cluster to benefit from the saved resources (bandwidth). It should be noted that the clustered users can receive the contents from the cluster head by using short range communication techniques like Wi-Fi Direct [71] and Device to Device (D2D).

Two conditions have to be satisfied before forming a user cluster. First, the users who request the same content are the ones who form the cluster. Second, the users should be or will be within a short range of each other. For that reason, the authors suggested using big data analytics to identify the users' closeness and to group the users into cluster(s). A cluster head is then selected among the nearby users, and the process is repeated among the base station users until there is either a cluster of users or a free (un-clustered) user(s). The simulation was carried for a single base station network and the results showed faster content delivery and improved throughput at the user level.

### 3.3.2 Improve cache node determination, allocation, and distribution accuracy in cognitive radio networks

In cognitive radio networks, Secondary Users (SU) have to leave the licensed spectrum when their activity starts to affect the QoS level of the licensed users. This move would require the existence of a cache node to compensate for the interrupted data transactions during the SU switch to the unlicensed spectrum.

The author of [72] proposed the use of big data analytics to process the data accumulated over time within the nodes. The goal was to utilize this data to reach a decision on the cache node distribution in a cluster network.

The author selected two out of three categories (open and selectively open systems) of cognitive radio networks. Due to the nature of the open systems, every SU willingly shares its information to be processed, which results in a large amount of data, so the prediction accuracy is high.

For the selectively open systems, the SU selectively shares its information with either some cache nodes, with the cluster head for a particular time interval, or with specific SUs in a cluster. This results in a variable amount of shared data, thus resulting in variable accuracy.

### 3.3.3 Tracking and caching popular data

The number of social network (i.e., Facebook and Twitter) users is massive. The multimedia contents of these networks are normally shared between common interest groups. However, big and important events attract a lot of attention and consequently a lot of content is shared across these networks. When a certain video or event goes viral, this sharing will eventually burden the network as the requested content would have to travel along the network on its way to the servers. The solution to such a problem was suggested by the authors of [68], they suggested monitoring popular and social media websites, analyzing the data, identifying if there is a growing interest in certain content, by which age category, and caching the popular data for a specific base station. Big data analytics can be of major use in this situation by employing it to do the required analysis. The result would be cached content available to the users faster (reduced provisioning delay) and without burdening the network.

### 3.3.4 Proactive caching in 5G networks

Cache-enabled base stations can serve cellular subscribers, this is done by predicting the most strategic contents and storing them in their cache. Thus, minimizing both the amount of time and the consumed network bandwidth, which can payoff in other ways (i.e., less congestion and less resource utilization).

An approach, proposed by the authors in [34], used big data analytics and machine learning to develop a proactive caching mechanism by predicting the popularity distribution of the content in 5G cellular networks. They demonstrated that this approach can achieve efficient utilization of network resources (backhaul offloading) and an enhanced user experience.

After collecting the raw data, i.e., the user traffic, the big data platform (Hadoop) has the task of predicting the user demands by extracting the useful information, like Location Area Code (LAC), Hyper Text Transfer Protocol (HTTP) request-Uniform Resource Identifier (URI), Tunnel Endpoint Identifier (TEID)-DATA, and TEID for control and data planes. Then using this information to evaluate the content popularity from the previously collected raw data. Experimentally testing this work on 16 base stations, as part of an operational cellular network, resulted in 100% request satisfaction and 98% backhaul offloading.

## 3.4 Network optimization

### 3.4.1 Big data-driven mobile network optimization framework

When thinking about optimizing a cellular network, it is important to collect as much information as possible. Large networks, as well as their users, generate a plethora of data, for which the use of big data analytics is vital to analyze the colossal amount.

The authors in [73] proposed a mobile network optimization framework that is Big Data Driven (BDD). This framework includes several stages, starting from the collection of big data, managing storage, performing data analytics, and the last stage of the process is the network optimization.

Three case studies were used to show that the proposed framework could be used for mobile network optimization.

**1- Managing resources in HetNets:**

The Mobile Network Operators (MNOs) may use big data to provide real time and history analysis across users, mobile networks, and service providers. MNOs can benefit from BDD approaches in the operation and deployment of their network, and this can be done in several stages:

A) **Network Planning**: Due to a deficiency in the level of sufficient statistical data, evolved Node B (eNB) sites are not optimally optimized, this can be dealt with if an adequate amount of information (user and network) is provided for analyses. Big data analytics can help MNOs reach better decisions concerning the deployment of eNB in the mobile network. The authors in [73] suggested the use of the network and anonymous users' data (e.g., dynamic position information and other service features). Providing a relation between the data and their events can offer a better understanding of the traffic trends. Big data sets provide actionable knowledge to reach an optimal decision concerning how and where to deploy eNBs in the network. Another important feature is the ability to prepare for future investments depending on the predicted traffic trends.

B) **Predictive Resource Allocation**: Resource requirements change depending on the density and usage patterns of mobile network subscribers. Predicting where and when mobile users are using the network can help in preparing for sudden significant traffic fluctuations. The authors in [73] suggested the use of big data analytics to examine behavioral and sentiment data from social networks and other sources. They also showed an interest in utilizing current and historical data to predict the traffic in highly populated areas within the network.

Using the cloud RAN architecture [74], the right place at the right time can be served through the predictive resource allocation, thus minimal service disruption can be achieved.

C) **Interference Coordination**: HetNets with small cells can be used to conduct interference coordination among macro and small cells. This coordination has to be carried out in the time domain instead of the frequency domain. Schemes like the enhanced Inter-Cell Interference Coordination (eICIC) in LTE-Advanced [75] efficiently enable resource allocation among interfering cells, as well as improving the inter-cell load balancing in the HetNets. eICIC allows Macro cells evolved Node B (MeNB) and its neighboring Small cell eNBs (SeNBs) to have data transmitted in isolated subframes, thus interference from MeNB to SeNB can be avoided. To implement eICIC, a special type of subframe named an Almost Blank Subframe (ABS) that carries minimum (and most essential) control information, was defined. It is worth noting that the ABS subframes are transmitted with reduced power [75], and that the network operator can control the configuration of that subframe.

Many factors contribute to the determination of the ABS ratio of the macro cell to the small cell, such as the traffic load in a specific area, the service type, and so on. The optimal ABS ratio varies dynamically, and this is due to the fact that inter-cell interference changes with time for the factors mentioned above.

In a BDD system, optimizing the radio resource allocation can be accomplished through the use of network analytics. The deployment of BDD optimization functions at the MeNB would enable them to collect and analyze eNB-originated raw big data (e.g., service characteristics and traffic features) in real-time, thus enabling a quick response. As a result, the performance optimization of each cell and the users can be fulfilled.

Optimizing ICIC parameters (e.g., ABS ratio) can be achieved by processing raw data in a periodic manner to acquire statistics and to detect traffic variations automatically.

Furthermore, the location and user traffic demands of multiple eNBs can be optimized, offering the deactivation of a SeNB due to elevated Signal-to-Interference-plus-Noise Ratio (SINR) to avoid the interference caused by a nearby SeNB that would also result in reducing the energy consumption.

**2- Deployment of cache server in mobile CDN**

Popular content (e.g., movies) can be delivered through a Content Delivery Network (CDN), which is a method that is considered efficient by many MNOs. Distributed cache servers should be located near the users to achieve a fast response as well as to reduce the delivery cost. In hierarchical CDN, it is vital to place cache servers in an optimal location. Due to the unique features that RAN has, it was the primary interest of the authors in [73].

It is expected that there will be an enhanced backhaul capability in 5G networks, and this would result in minimizing the concerns related to the latency and traffic load of backhaul transmissions. Therefore, not all MeNBs would require a dedicated distributed cache server. In addition, a SeNB can have a distributed cache server.

Optimal cache server placement depends on several factors, such as the features and load of traffic in a given area, as well as the cost of storage and streaming equipment. To help the MNOs decide where to deploy their cache servers, data analytics methods can be regarded as a feasible solution. However, this would require the collection of all the above-mentioned factors over a long period in the related coverage area.

**3- QoE modelling for the support of network optimization:**

The authors of [73] believed that the management of services and applications needed more than just relying on the QoS parameters. Instead, they suggested taking the quality (i.e., QoE), as perceived by the end users, to be regarded as the optimization objective. Accurate and automatic real time QoE estimation is important to realize the optimization objective. In addition to the technical factors, non-technical factors (e.g., user emotions, habit, and expectations, etc.) can affect the QoE. A profile for each particular user comprising the above non-technical factors would help in the QoE evaluation.

Since answering the questions that would lead to a clear profile is not a task that would be fancied by a typical user, the authors suggested installing a profile collection engine on the users' mobile devices. User activities are compared and tracked to recognize differences and similarities, and then they are stored in a database for additional processing. After profiling, the following step constitutes the use of machine learning to identify the relationship between QoE and the influencing factors.

Data analytics can be used to discover what impacts the QoE in users' devices, as well as the services and network resources. The next step is for network optimization functions to react to determining what caused the problem and select the optimal action accordingly.

### 3.4.2 Improve QoS in cellular networks through self-configured cells and self-optimized handover

Cellular networks have a crucial element on which the concept of mobility depends. This element is the handover success rate, which ensures call continuity while the user moves from one cell to another. Failing in that particular element would impact the quality of the service, thus putting the operator into a questionable situation.

Operators try to make sure that each cell has a list of manually configured and optimized neighbor cells. Hence, it is vital to note the high probability of these cells failing to adapt when a rapid response is required due to a sudden network change.

The authors in [76] presented two methods that used big data analytics to introduce a self-configured and self-optimized handover process, the first was associated with newly introduced cells, while the latter was concerned with the already existing cells. The analysis started by collecting and archiving predefined handover KPIs. A dispatcher process is run after the collection period, and its aim is to check the files to see if they were marked as new cells (where Self-Configuration Analytics is started) or not (where Self-Optimization Analytics is started):

**1- NCL self-configuration for new cells:**

Newly installed base stations require Neighbor Cell List (NCL) to be configured on the new cells. The selection process takes into consideration the antenna type, the azimuth angle (for directional cells), the geographic location of the candidate cells, and the process concludes by selecting cells with a minimum distance and maximum traffic load to be the top candidate cells. The NCL is configured via Network Management System (NMS) Configuration Management (CM) tools.

**2- NCL self-optimization for existing cells:**

The process starts by collecting KPI measurement statistics for the failed and successful handovers, and this task is done by the Performance Management (PM) or the NMS.

Cells with a handover failure rate below a predefined threshold are excluded from the NCL, while unlisted neighboring cells with a successful rate above a predefined threshold are considered as new neighboring cells.

### 3.4.3 Optimizing the resource allocation in LTE-A/5G networks

The overall system performance evaluation in advanced wireless systems, like LTE, depends on KPIs. In a quest to enhance the user experience, the authors of [77] proposed an approach that utilizes user and network data, such as configuration and log files, alarms, and database entries/updates. This approach relies on the use of big data analytics to process the above-mentioned data. The ultimate goal is to provide an optimal solution to the problem of allocating radio resources to RAN users, and guarantee a minimal latency between requesting the resource and allocating it. This is done through user and network behavior identification, which is a task well-matched for big data analytics.

The proposed framework involves three stages:

*First stage*: This process is carried out in the eNB system, processing the data from the cellular and core network side. Binary values are acquired by comparing cellular level KPIs to their respective threshold values, thus keeping the binary matrix updated. This procedure is repeated at fixed intervals.

*Second stage*: Repeat the same steps as in the first stage. However, this process is carried out on subscriber level data to acquire subscriber KPI, and maintain a binary matrix.

*Third stage*: This is activated when a user initiates a resource allocation request. A binary pattern is generated based on the user requirements. This pattern is later handed over to stage 2 to update the binary matrix (if required) and incorporate the new values in the row that represents the requested bandwidth. After generating the updated row, it is transferred to the first stage for comparisons with the current Physical Resource Block (PRB) groups. To identify which PRBs suit the user, the fuzzy binary pattern-matching algorithm [78] was used for that purpose.

Using this algorithm, the execution time increased linearly for an exponential increase in the number of comparison patterns.

### 3.4.4 Framework development for big data empowered SON for 5G

The authors of [79] proposed a framework called Big data empowered SON (BSON) for 5G cellular networks. Developing an end-to-end network visibility is the core idea of BSON. This is realized by employing appropriate machine learning tools to obtain intelligence from big data.

According to the authors, what makes BSON distinct from SON are three main features:

- Having complete intelligence on the status of the current network.
- Having the ability to predict user behavior.
- Having the ability to link between network response and network parameters.

The proposed framework contains operational and functional blocks, and it involves the following steps:

1- *Data Gathering*: An aggregate data set is formed from all the information sources in the network (e.g., subscriber, cell, and core network levels).

2- *Data Transformation*: This involves transforming the big data to the right data. This process has several steps, starting from:

  a. *Classifying* the data according to key Operational and Business Objectives (OBO), such as accessibility, retainability, integrity, mobility, and business intelligence.

  b. *Unify/diffuse* stage, and the result of this stage is more significant KPIs, which are obtained by unifying multiple Performance Indicators (PIs).

  c. According to the KPI impact on each OBO, the KPIs are *ranked*.

  d. *Filtration* is performed on the KPIs impacting the OBO less than a pre-defined threshold.

  e. *Relate*, for each KPI and find the Network Parameter (NP) that affects it.

  f. *Order* the associated NP for each KPI according to their association strength.

g. Cross-correlate each NP by finding a vector that quantifies its association with each KPI.

3- *Model*: Learn from the right data acquired in step 2 that will contribute to the development of a network behavior model.

4- *Run SON engine*: New NPs are determined and new KPIs are identified using the SON engine on the model.

5- *Validate*: If a new NP can be evaluated by expert knowledge or previous operator-experience, proceed with the changes. Otherwise, the network simulated behavior for new NPs is determined. If the simulated behavior tallies with the KPIs, proceed with the new NPs.

6- *Relearn/improve*: If the validation in step 5 was unsuccessful, feedback to the *concept drift* [80] block, which will update the behavior model. To maintain model accuracy, concept drift can be triggered periodically even if there was a positive outcome in the validation step.

### 3.4.5 User-centric 5G access network design

Enhancing the user experience, giving a higher data rate, and reducing the latency are considered the key goals of a 5G system. The authors of [27] expect the following to be the key elements in the design of a user-centric 5G access network.

#### 3.4.5.1 Personalized local content provisioning

It is important for the access network to evolve from being user and service agnostic, by acting merely as a blind pipe that connects the user to the core network, to becoming user-centric. The latter term would direct the network access to the right path of being user and service aware, thus facilitating a key technology in 5G, which is local content provisioning, this in-turn would pay off in the form of end-to-end latency reduction and enhancing the user experience.

The role of big data analytics becomes very clear, as it provides a necessary ability of predicting user requirements. These requirements can be met in the case of local availability.

The authors of [27] proposed several steps to achieve content provisioning, and they are as follows:

- *Traffic and user Information Acquisition*: Traffic attributes (application type, server address, and port number, etc.) are collected through packet analysis, and analyzed using a clustering algorithm (e.g., k-means [81]) to perform traffic labelling (news, sports, and romance, etc.).

- *Analyzing and Predicting User Requirements*: This step is achieved by a big data analysis algorithm (an algorithm like collaborative filtering was recommended [82]), which utilizes the traffic attributes mentioned above, and thus recommends content based on user's interests.

- *Local Caching and Content Management*: Popular content is provided in the form of local copies. Content that might be of interest to the users are to be locally cached after being downloaded from the application server.

- *Content Provisioning*: When an application request is initiated by the user, the system will check if the content is already locally cached, so it can be sent directly. Furthermore, big data analysis can give content recommendations, thus the system will check if these contents are cached locally, and push them directly to the user.

#### 3.4.5.2 Providing flexible network and functionality deployment

The authors of [27] noted that processing the characteristics of both the regional user and service using big data analytics can be of great help in flexible network and functionality deployment in the following ways:

- Flexible Network Deployment
  Since 5G will support diverse low-cost Access Points (APs), such as coverage APs and hot spot APs, using big data analytics can be useful to forecast the traffic characteristic, hence establish the base for achieving a dynamic network deployment for the APs.

- Flexible Functionality Deployment
  Analyzing and predicting regional user and service requirements can be realized using big data analytics. This will form the foundation of dynamic functionality deployment, which will help decide where to deploy certain functionality modules (e.g., safety modules where there are security requirements).

#### 3.4.5.3 Using user behavior awareness to optimize wireless resources 5G networks

Optimization of wireless resources should be carried-out according to the user and 5G service requirements. This is done to enhance the user experience and improve the efficiency.

According to what the authors of [27] proposed, big data can be used to analyze user mobility patterns, predict the motion trajectory, and hence pre-configure the network accordingly. Each user's historical Access Point (AP) list is recorded by the APs. This data can be uploaded to a central module for processing, or to a target AP in case the service AP was altered. Using a big data analytics algorithm, the collected data is analyzed to forecast the motion trajectories.

#### 3.4.5.4 Big data-based network operation system

The authors of [27] proposed a system that can maximize the efficiency of big data based network operation. This can be fulfilled by optimally allocating the network resources to each AP and user. The proposed system consists of two parts:

1- Decision making domain

This is responsible for collecting and managing the user, network upgrades, configuration, service, and terminal state information. This domain exploits big data analytics to provide the basic configuration essential for network initialization. For this domain to function properly, it would need to realize the whole profile of the user and service requirements, as well as the functionality distribution in the network.

2- Implementation domain

Its responsibility is mainly for status reporting of the user, terminal, and network, dynamic deployment, and network configuration. Depending on the requirements after acquiring the personalized configuration, this domain can use the dynamic APs functionality and configuration to build multi-connectivity bearers with terminals.

### 3.4.5.5 Multi-RAT or HetNet energy saving

Small cells are used in multi-Radio Access Technology (RAT) or HetNet mobile networks. For energy saving purposes, when traffic falls below a certain threshold, small cells may enter into a dormant state, and this forces the small neighboring cells to serve the dormant cell's coverage area. Several energy saving schemes were discussed in [83]. However, these schemes failed to adapt to the dynamic temporal and spatial traffic variations, as they operate under a relatively long time scale leading to unacceptable delay. This happens because when a number of newly arrived User Equipment (UE) seek access to the network, where cell activation is essential, the corresponding small cells requires to be first activated, before operating normally. Only at that point can the UEs access the network using the standard procedure. UEs, however, would suffer from an unacceptable delay when trying to access the cells.

The authors of [27] proposed a user-centric approach that is based on big data analytics. The proposed solution claims to achieve optimal implementation for the activation of small cells as well as UE access to the network. Thus, joint optimization for both UEs' access and energy saving can be achieved.

### 3.4.6 Network flexibility using consumption prediction

Consumption analysis is concerned with two factors: customer locations and type of service. Consumption trends can be classified in a timely manner into long-term, seasonal, and short-term.

To reach an accurate prediction, the authors in [68] implied that user data (e.g., GPS location and service usage) can be correlated with other data (e.g., news, social network, events, and weather conditions). Using big data analytics to analyze these correlations, operators would be able to decide when and where to place their nodes without affecting the subscribers' satisfaction.

## 4. The role of big data analytics in SDN & intra-data center networks

SDN offers the ability to program the network with a centralized controller, this controller is capable of programming several data planes using one standardized open interface, thus providing flexible architectural support [1]. The following research topics utilized the properties of both Software Defined Network (SDN) and big data analytics by employing the analysis results to program the network. Those topics can be classified according to the area under discussion as follows:

1- Traffic Prediction: The paper surveyed in this section employ traffic prediction to optimize network resource allocation.

2- Traffic reduction: Pushing the aggregation from the edge towards the network.

### 4.1 Traffic Prediction

#### 4.1.1 Cognitive routing resources in SDN networks

The network's next generation has to be smart and flexible, with the ability to modify its strategy according to the network status in an automatic manner. To simplify the network management, SDN has made the above requirement possible by decoupling the control and forwarding planes through the OpenFlow protocol [84].

OpenFlow is considered to be the first standardized protocol in SDN, it is also identified as the enabler of SDN. SDN/OpenFlow has influenced Google to switch to OpenFlow in its inter-data center network, which resulted in an approximate 99% increase in the average Google WAN link utilization [85].

The architecture proposed by the authors of [85] included the following parts:

**1- User preference analysis server:**

The authors adopted the Hadoop platform to realize the prediction functionality. They utilized the analyses of both network traffic and user application information to find each application flow distribution. For each data flow, they found a specific distribution law. They analyzed that law, and for different applications and areas they developed a preliminary general prediction model to fit the case of the same application but in different areas.

**2- Interface design between SDN controller and database:**

A cloud platform is responsible for calculating and predicting the flow distribution values of each OpenFlow switch. In addition, this platform will read the link information and perform traffic prediction. A database will hold the recorded values and the last predicted values will be updated. To ensure that the allocation of resources accommodates the traffic variation, Floodlight (a Java-based SDN controller that can accommodate different applications by loading different modules) will read the newest predicted values from the database regularly.

**3- SDN controller-based routing module:**

The predicted values are used as preferences to select the best route. The researchers used an improvement on the Dijkstra algorithm.

Application awareness and the prediction of user preferences were integrated into SDN through the newly proposed architecture, which would facilitate and enhance network resources allocation and provide better application classification.

The role of big data is exemplified by its ability to use network flow analyses and users' behavior to forecast the type and rate of the incoming traffic flows.

The procedure proposed by the authors of [85] is as follows:

1- Current network load (i.e., traffic volume and type, etc.) is read by a cloud platform.

2- The overall traffic is predicted in advance and gathered in a database. This is done using a big data-powered prediction algorithm.

3- The SDN controller accesses the database and reads the stored data mentioned in step two.

4- A resource allocation scheme is created by the SDN controller using the above-mentioned routing algorithm, and this scheme is sent later to the related switches.

Big data analytics can use the users' requirements to provide the network with a dynamic resource allocation and application classification, hence providing the network with better load balancing techniques.

The results from the implemented testbed showed the ability of the proposed solution to self-adapt towards flow variation by dynamically issuing traffic tables to the related switches, which can increase the resource utilization and attain an improved overall load balance.

### 4.1.2 Predicting data communication volume at runtime in data center networks using SDN

Networks that deal with big data applications may suffer from the size and speed of data. For example, the networks' overall response time is affected by MapReduce's heavy-communication phase. This problem can be intensified if the communication patterns experience a heavy skew impact. The authors of [86] have proposed Pythia, which is a system that can optimize the data center network at the runtime by utilizing the real time communication prediction of Hadoop. It also maps the end-to-end flows to the underlying network.

Pythia utilizes the SDN-offered programmability to achieve efficient and timely network resource allocation for shuffle transfers.

Depending on the network workload and blocking ratio, the Hadoop workload saw a consistent acceleration when using Pythia, and job completion time was reduced between 3% and 46% in comparison with MapReduce benchmarks.

### 4.2 Traffic Reduction

### 4.2.1 Increasing network performance through traffic reduction

Large networks, such as those of Google and Facebook, or even small and medium sized enterprise networks suffer from a plethora of traffic. This happens due to the colossal amount of data being processed either in batch or real time applications.

A common solution is to increase the available bandwidth in the enterprise clusters. However, the authors of [87] proposed another approach that improves the network performance, pushing the data aggregation from the edge towards the network, thus decreasing the traffic.

A platform called CamCube [88] was used; it substitutes the use of dedicated switches by distributing the functionality of the switch across the servers. It is worth noting that CamCube offers the ability to intercept and modify packets at each hop. In addition, it uses a direct-connect topology, in which, a 1Gbps Ethernet cross-over cable is used to connect servers to each other in a direct way.

Exploiting CamCube's properties to realize high performance, Camdoop, which is a CamCube service that runs MapReduce-like jobs, was used. It offers full on-path data stream aggregation. Camdoop builds aggregation trees where the children are resembled by the intermediate data sources, while the roots are located at the servers performing the final reduction in traffic.

A small prototype of Camdoop running on CamCube was tested, and a simulation was used to show that the same properties still hold at scale. The results showed a significant traffic reduction with the proposed system when compared to Camdoop running on a switch and when compared to systems like Hadoop and Dryad/DryadLINQ [56] [89].

## 5. The Role of Big Data Analytics in Optical Networks:

This section discusses research papers that employ big data analytics for optical network design, the topics are classified as follows:

1- Network optimization: Here the parallel processing merits of Hadoop are utilized to reduce the execution time of different (bin-packing) optimization algorithms.
2- Traffic Prediction: Using big data analytics to dynamically reconfigure the network according to predicted traffic.

### 5.1 Providing solutions to network optimization problems

### 5.1.1 Solving the RWA problem

The Routing and Wavelength Assignment (RWA) algorithm [90] plays an important role in optical networks. The authors in [91] considered the RWA algorithm to be an illustration of the bin-packing problem that is listed as a classical NP-hard problem [92].

The authors in [91] used a Hadoop cloud computing system that consisted of 10 low-end desktop computers to independently run an instance of the RWA algorithm on each of them for a certain number of demand sequences. The goal is to sufficiently evaluate the demand sequence within a short period. The procedure is as follows:

1- An input file is fed to the HDFS, it incorporates the information of the lightpath demand requests.
2- The file is read by the map function, where the demand list is regarded as a value and combined with different keys ranging from 0 to 19 that later serve as random seeds in the reduce functions. It is worth noting that the authors set two reduce functions per computer (i.e., a total of 20 reduce functions).
3- The key-value pairs are then forwarded to the 20 reduce functions where parallel computing is conducted. The lightpath demand list is shuffled 250 times in a random fashion (i.e., 5000 shuffled demand sequences), and this happens for each key-value pair within each reduce function. To acquire the number of required wavelengths, the RWA heuristic is run for each of the shuffled demand sequences. The optimal result of each reduce function is then compared against the remaining 19 to find the global optimum.

Different test networks (ranging from 20 to 500 nodes) were used to evaluate the performance of the Hadoop system. The results were optimality evaluated by comparing them against the results of an Integer Linear Programming (ILP) optimization model and they showed a close proximity to the optima (except for two cases). It is worth noting that the ILP approach assumes full wavelength conversion, which plays the role of the performance's lower bound in the present evaluation.

### 5.1.2 Solving multiple optimization problems using Hadoop

The authors in [93] proposed to solve several optimization problems in the optical network paradigm. The problems are:

1- Energy minimization problem [94], where the goal is to minimize the overall network power consumption from non-renewable energy sources.

2- Shared Backup Path Protection (SBPP) –based elastic optical network planning problem [95], where a heuristic used the concept of Spectrum Windows Planes (SWPs). The objective was to minimize the maximum number of Frequency Slots (FSs) in the network.

3- Adaptive Forward Error Correction (FEC) assignment problem [96], where the goal is to maximize the total number of FSs utilized for user data transmission. A heuristic based on SWPs was developed to solve the Routing and Spectrum Allocation (RSA) problem.

The above problems are of a bin-packing type and classified as NP-hard. Several aspects (i.e. demand size and route) should be considered when serving network traffic demands. Due to the high computational complexity and the order of served demands, the performance of heuristic algorithms trying to solve these problems cannot be guaranteed. This is because of using the simple largest to smallest ordering strategy. Good performance can be attained by randomly shuffling demand sequences, then implement a heuristic algorithm for each sequence and choosing the one with the optimum performance. To shorten the computation time and to overcome the computational complexity, a Hadoop cloud computing system consisting of seven computers was proposed by the authors in [93]. This way, a heuristic algorithm can be executed for multiple shuffled demand sequences in a parallel manner.

The Hadoop MapReduce platform makes it possible to evaluate multiple shuffled demand sequences in parallel. A heuristic algorithm serves each of the shuffled demand sequences and a result is produced each time. The results are then compared and the one with the best performance is chosen. The same procedure is repeated on each Reduce function. The final global optimum is found by comparing the results across all reduce functions.

Performance evaluation is done by employing two test networks; the 24-node, 43-link USNET network (adopted for problems 1 and 3) and the 11-node, 26-link COST239 network (used for problem 2). For the first optimization problem, the total consumption of non-renewable energy decreased by 8% (when the number of shuffled demand sequences increased from 1000 to 10,000). As for the second optimization problem, the number of required FSs is significantly reduced. In the third problem, the total number of FSs for user data transmission was increased and 3% performance improvement was noted when compared with the case of running Hadoop on a single machine. The computation time for all three problems was significantly shorter compared to a single-Hadoop machine.

### 5.2 Dynamic network reconfiguration

#### 5.2.1 VNT adaptability using traffic prediction

The emergence of new services is placing new demands on networks in terms of large and dynamic bit rate requirements. This caused network operators to look for a Virtual Network Topology (VNT) architecture that can cope with the anticipated traffic in a dynamic fashion. One solution is realized by overprovision the network, however, the downside is the increase in Total Cost of Ownership (TCO). Another solution saves power by using threshold-based capacity reconfiguration [97]. The drawback is that there is no saving in the number of transponders that needs to be installed in each IP router compared to overprovisioning.

An alternative solution is proposed in [59] where VNT reconfiguration can be attained regularly using big data analytics. This is done by periodically analyzing Origin-Destination (OD) traffic so that VNT reconfiguration can be performed accordingly. Collection of traffic monitoring data takes place at edge IP routers regularly. A set of traffic samples is collected by every edge router for every other destination router. These sets are stored in a *collected data repository*. According to a predefined time period, the collected data is then summarized for each OD pair by periodically retrieving the collected data repository and performing data stream mining sketches. The result of this stage is a *modeled data repository* which includes, among others, maximum, average, and minimum bit rate for every OD pair. Using machine learning techniques (i.e. Artificial Neural Networks), a *prediction module* generates the predicted OD traffic matrix for the upcoming period. The decision on whether to perform VNT reconfiguration or not is determined by a *decision-maker* module by relying on the above-mentioned matrix. If a reconfiguration is required, a *VNT optimizer* is provided with both current and predicted OD traffic matrices. The solution is fedback to the network controller to implement the required changes in the data plane.

The performance of the proposed solution was compared against the static and the threshold-based methods. An overall saving between 8% and 42% in the number of installed transponders was achieved. The proposed solution has the ability to react during low traffic hours by deactivating transponders, thus, energy saving is attained. Also, the advantage of cost reduction is attained by releasing lightpaths from the underlying optical layer.

## 6. The role of big data analytics in network security

### 6.1 Peer-to-Peer Botnet detection

Many security problems on the Internet are caused by Botnets, which can be defined as networks of malware-infected machines controlled by an adversary [98]. Botnet attacks form a real security concern, with the ability to utilize 90,000 IPs in an attack [99], it is a challenge on an international scale, especially when taking into consideration the financial damage they can inflict.

To detect and neutralize such attacks, security researchers and network analysts consider packet capturing and network tracing to be amongst the most appreciated resources. However, analyzing these massive sized datasets is not an easy task for today's computers. To overcome that challenge, the authors of [100] proposed a scalable threat detection framework that uses the following components:

1- *Traffic sniffer*: Dumpcap [101] is used for packet sniffing while Tshark [101] is used for field extraction, and the fields are then submitted to the HDFS for storage.

2- *Feature Extraction Module*: The HDFS-submitted files are then processed by Apache Hive [102] for feature extraction.

3- *Machine Learning Module*: Scalability is a requirement when it comes to the machine learning module. This requirement is met using a machine learning library called Mahout [103], thus harnessing the cluster high computational power to achieve optimized results. It is worth noting that Mahout is built on top of Hadoop, and its classification and clustering core algorithms are run as MapReduce jobs.

The proposed approach achieved a detection time within tens of seconds, and the authors claimed that this time can be reduced to less than 10 seconds after some Hadoop tweaking and additions to the cluster.

### 6.2 *Improving network security by discovering multi-pronged attacks*

Networks are considered a target for intruders who would try to infiltrate them. Multi-pronged attacks may spread over network subnets; the spreading might target several scattered network points or take place in different events over time.

To discover and predict such attacks, the authors of [104] proposed a system named Big-distributed Intrusion Detection System (B-dIDS) that relies on two components:

1- HAMR: An in-memory MapReduce engine used for big data processing. It is worth noting that HAMR supports both batch and streaming analytics in a seamless manner.

2- An analytics engine: Residing on top of HAMR, the analytics engine includes a novel ensemble algorithm. Its basic principle relies on using clusters with multiple IDS alarms to extract the training data.

The proposed system scans the IDS log data, checking for alarms that might be treated as unthreatening at first glance (when examined separately) but that may result in an opposite judgment when combining them with other alerts.

### 6.3 *Device fingerprinting in wireless networks*

Big data analytics and machine learning have several algorithms in common [105]. More security concerns can be addressed through the use of machine learning algorithms such as device fingerprinting. The authors in [106] have conducted a survey on wireless device fingerprinting methods in wireless networks. They illustrated the main features and techniques used towards this end. Device fingerprinting can be defined as the process of generating device-specific signatures by gathering device information. This is done through analyzing the information across the protocol stack, and it can be used to counter the vulnerability of wireless networks to insider attacks and node forgery. Two types of fingerprinting algorithms were discussed; white-list based (i.e. supervised learning) and unsupervised learning based approaches.

The device fingerprinting process is broken into three main steps; *step one* is concerned with identifying relevant features found in all layers across the protocol stack. *Step two* is where features are extracted and modeled. The features tend to be stochastic in nature due to the dynamic nature of wireless channels, consequently, the models will be stochastic as well. *Step three* is where device identification takes place by employing machine learning algorithms (supervised and unsupervised).

The authors reviewed the existing algorithms and concluded that despite the high computational complexity of unsupervised learning methods, their role is limited to detect the presence and the likely culprit involved in the attack while failing to pinpoint the malicious devices in an exact manner. In spite of the limitation, unsupervised learning approaches showed further practicality when compared to white-list based approaches, as they require no pre-registration process and human intervention.

### 6.4 *Machine learning methods for cybersecurity intrusion detection*

The authors in [107] surveyed the topic of intrusion detection methods based on data mining and machine learning algorithms. They compared different methods taking into account complexity, accuracy, understandability of the final solution, and classification time of an unknown instance using a trained model. They referred to the availability of labeled data as the biggest gap that, if bridged, can lead to significant advances in machine learning and data mining methods in the field of cyber security. They also highlighted an open area for research, namely the investigation of fast incremental learning methods to update misuse and anomaly detection models on a daily bases.

Finally, we summarise the research outcomes related to big data analytics-based network design in **Fig 3**. It is clear that the wireless field is getting most of the researchers' attention compared to the other fields. This may be attributed to the more significant challenges faced by wireless networks compared to wired networks and hence the more significant level of opportunities. The larger numbers of papers addressing the use of big data analytics and methods in wireless may also be a reflection of the larger number of researchers that focus on wireless networks compared to wired networks. Furthermore, we present a summary in **Table 2** for all the research topics illustrated between sections 2 and 5.

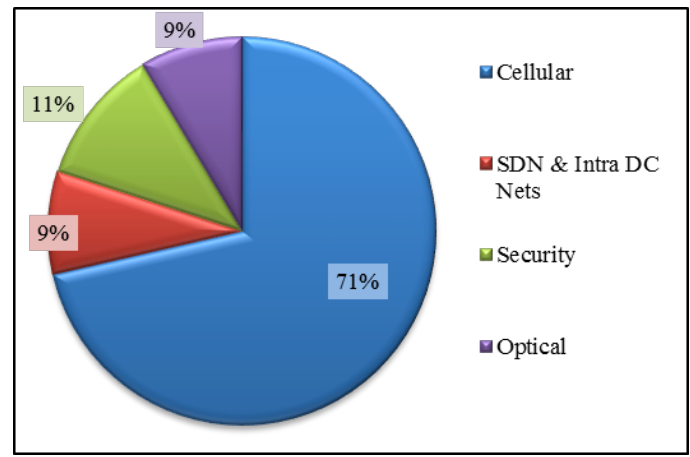

**Fig. 3.** Percentage of surveyed research topics according to subject area.

**Table 2: Research summary**

| Network Type | Research Category | Reference | Proposed or Deployed Technique |
|---|---|---|---|
| Wireless | Failure Prediction, Detection, Recovery, and Prevention | [63] | Analyzed inter-technology (2G-3G) failed handovers. |
| | | [27] | Used XDR data to discover network failures and present a solution advice. |
| | | [64] | Developed CADM which uses CDRs to identify anomalous sites. |
| | | [65] | Presented three case studies of self-healing using big data analytics. |
| | | [68] | Suggested the analysis of the bandwidth trends to predict equipment failure. |
| | Network Monitoring | [69] | Developed a Hadoop-based system to monitor and analyze network traffic. |
| | | [70] | Developed a solution powered by big data platforms with distributed storage and distributed database to solve the issues of data analysis and acquisition. |
| | Cache and Content Delivery | [26] | Utilized big data to form a cluster made up of nearby users that share the base station's wireless channel. |
| | | [72] | Analyzed the data that resides within the cache nodes to enhance the determination, allocation, and distribution of cache nodes. |
| | | [68] | Suggested monitoring and analyzing social media and popular sites, to predict and cache certain contents, according to age category, at the predicted locations where these contents are highly demanded. |
| | | [34] | Proposed the use of big data analytics and machine learning techniques to proactively cache popular content in 5G networks. |
| | Network Optimization | [73] | Presented three case studies in which a proposed network optimization framework is efficiently utilized. In particular, the work suggested:<br>1) The use of big data analytics to manage resources in HetNets. This is done in three stages (network planning, resource allocation, and interference coordination).<br>2) The deployment of cache servers in mobile CDN.<br>3) The optimization of networks with QoE in mind. |
| | | [76] | Proposed NCL self-configuration/optimization algorithms to achieve an automatic, self-optimized handover. The work relied on the processing of CM and PM KPIs using big data analytics platform. |
| | | [77] | Developed a three-stage framework that utilizes the network and user KPIs to reach an optimal allocation of radio resources (PRBs). |
| | | [60] | Presented a framework that uses big data collected from the cellular network to empower SON. They also presented a case study on how to detect sleeping cells using this framework. |
| | | [27] | Investigated the impact of big data on 5G networks in terms of:<br>1) Efficient content provisioning.<br>2) Flexibility in functionality and network deployment.<br>3) Utilizing user behavior in wireless resource optimization.<br>4) Achieving highly efficient network operation.<br>5) Saving energy in HetNet or Multi-RAT networks. |
| | | [68] | Correlated location data, service usage, and other contextual data to predict the consumption trends and select the optimal node location. |
| SDN and Inter-Data Center | Traffic Prediction | [85] | Dynamic allocation of network resources by relying on traffic predicted by employing Hadoop platform. |
| | | [86] | Developed Pythia, a system that uses Hadoop's properties to predict the volume of data communication at runtime in data center networks. |
| | Traffic Reduction | [87] | Proposed Camdoop and run it over CamCube, the performance surpassed that of Camdoop running on a conventional switch. |
| Optical | Network Optimization | [91] | Used Hadoop to find a solution for the RWA problem. |
| | | [59] | Predicted future traffic by using Big Data Analytics to reconfigure VNT regularly. |
| | Traffic Prediction | [93] | Employed Hadoop MapReduce to solve multiple optimization problems of bin-packing nature. |
| | Security | [100] | Proposed a threat detection framework to detect peer-to-peer Botnet attacks. |
| | | [104] | Developed B-dIDS, a system that scans IDS log files to detect multi-pronged attacks in distributed networks. |
| | | [106] | Surveyed the topic of wireless device fingerprinting methods and how machine learning algorithms can be used for device identification. |
| | | [107] | Provided a survey discussing the role of data mining and machine learning algorithms in intrusion detection methods. |

## 7. Big data analytics in the industry

Throughout our survey, we came across several companies that offer network solutions based on big data analytics. These companies and solutions are highlighted in **Table 3**. It should be noted that these solutions are enabled by research conducted in their corresponding areas. We have added academic research papers related to each solution in Table 3.

Due to the proprietary nature of industrial products, the exact algorithms or methods behind these products is not available in the open literature. Therefore, academic papers with related concept(s) are cited. *NetReflex IP* and *NetReflex MPLS* utilizes big data analytics [27, 73, 108] to provide services like anomaly analysis and traffic analysis. Nokia provided several solutions targeting the wireless field. For example, *Traffica* introduces itself as a real-time traffic monitoring tool that analyzes user behavior to gain network insights, similar approaches were presented in academia by the authors of [69, 109]. The *Wireless Network Guardian* detects user anomalies in mobile networks where a comparable topic was discussed in [110]. *Preventive Complaint Analysis* makes use of big data analytics to detect behavioral anomalies in network elements where the authors in [111] provided a similar approach. *Predictive Care* utilizes big data analytics to identify anomalies in network elements before affecting the user, a comparable academic approach is presented in [110, 112]. HP presented *Vertica*, a solution that exploits CDRs for network planning, optimization, and fault prediction purposes.

**Table 3**: Big data analytics-powered industrial solutions.

| No. | Manufacturer | Solution Name | Related Academic Papers | Usage, Functions and Capabilities |
|---|---|---|---|---|
| 1 | Juniper | NetReflex IP | [27, 73, 108] | Eliminates network errors. |
| | | | | Monitors QoS/QoE. |
| | | | | Capacity planning, traffic routing, caching, and other optimizations. |
| | | NetReflex MPLS | | Segment and trend MPLS and VPN usage to plan for congestion. |
| | | | | Identifies traffic utilization and trends to optimize operational cost. |
| | | | | Ability to slice network performance according to VPN, Cost of Service (CoS), and Provider Edge (PE)-PE enabling more efficient planning. |
| 2 | Nokia | Traffica | [69, 109] | Real-time issues detection and network troubleshooting. |
| | | | | Gain real-time, end-to-end insight on traffic, network, devices, and subscribers. |
| | | Wireless Network Guardian | [110] | Improves end-to-end network analytics and reporting with real-time subscriber-level information. |
| | | | | Detects anomalies and reports airtime, signaling, and bandwidth resource consumption. |
| | | | | Proactive detection of issues, including automatic detection of user anomalies and low QoE score alerts. |
| | | Preventive Complaint Analysis | [111] | Detects network elements' behavior anomalies. |
| | | | | Predicting where customer complaints might arise and prioritizes network optimization accordingly. |
| | | Predictive Care | [110, 112] | Used for network elements, and proved its effectiveness by helping Shanghai Mobile become more agile and responsive. |
| | | | | Accuracy of the simplified alerts is around 98 percent, reducing operational workload. |
| 3 | HP (HPE) | Vertica | [64, 113] | Provides CDR analysis that can help Communication Service Provides (CSPs). |
| | | | | Examines dropped call records and other maintenance data to determine where to invest in infrastructure. |
| | | | | Failure prediction and proactive maintenance. |
| 4 | Amdocs | Deep Network Analytics | [114] | Combines RAN information with BSS and customer data to deploy the network proactively. |
| | | | | Predictive maintenance. |
| 5 | Apervi | Apervi's Real-time Log Analytics Solution (ARLAS) | [115-117] | Collects, aggregates, and stores log data in real-time. |

The authors in [64, 113] researched akin approaches. Amdoc's *Deep Network Analyzer* provides predictive maintenance and proactive network deployment for cellular networks. The authors in [114] presented a similar approach. Log analytics can be used for a variety of purposes. Aprevi's *ARLAS* solution provided real-time collection and storage of network logs. Related academic research was presented by the authors in [115-117]. Examining the above solutions, one can note that the majority of the solutions are in the wireless field. This, in fact, coincides with the orientation of the academically-researched topics. Sampling through the offered solutions, we noticed the increased interest in anomaly prediction and network node deployment. Thus, offering the customer a service that is as close to optimal as possible, while minimizing network expansion expenditures.

## 8. Big data analytics-powered design cycle and challenges

In this section, we are highlighting a common theme among most of the surveyed papers. This can be realized as depicted in **Fig. 4**. Also, we are illustrating the challenges facing the implementation of big data analytics in network design and operation.

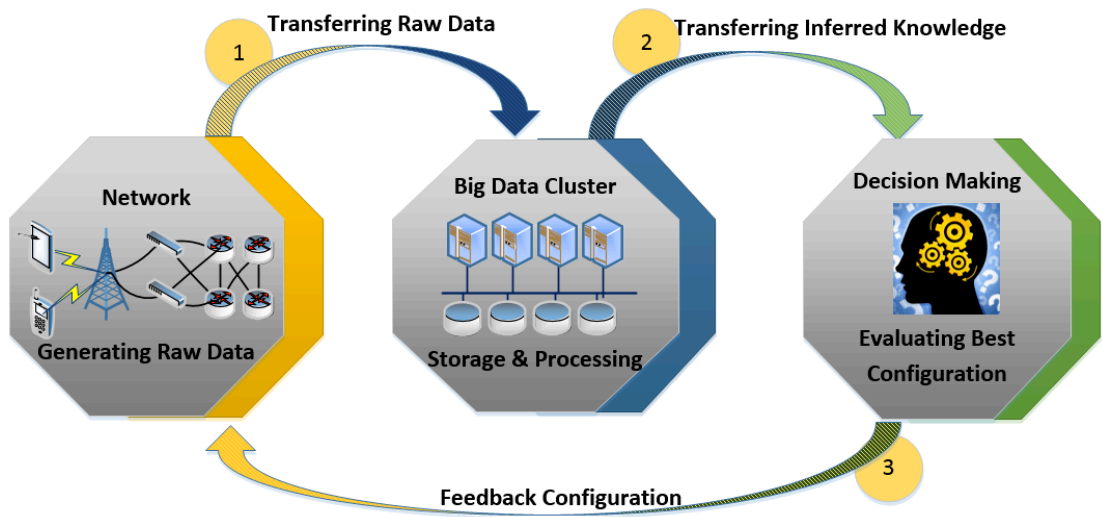

**Fig. 4.** Big-data-powered network design cycle.

## 8.1 Big data analytics design cycle

The quest for a well-designed communication network is never-ending. Researchers in the big data era rely on the capabilities offered by big data analytics to transform the way networks are being designed. This includes employing big data analytics to predict and minimize the bandwidth utilization, anticipate and prepare for upcoming failures, and predict the precise energy requirements. Hence, creating a network with fewer outages, higher user satisfaction, and an enhanced performance.

The network design process using big data can be outlined as shown in **Fig. 4.** Big data is collected from the network, stored, and processed in a big data cluster to extract useful information, such as trends, patterns, and correlations (step 1). The resulting information is then transferred to the decision-making platforms where a new design decision for the network is evaluated by algorithms based on the inward inferred knowledge (step 2). Finally, the new design decision is sent as feedback configuration parameters to the network where re-configuration is implemented (step 3).

It should be noted that the duration of the above-mentioned cycle might vary depending on the application type of the network, e.g., enterprise, healthcare, agriculture, or transportation. For instance, enterprise networks can generate large amounts of data over a short period and usually configuration faults could be undone anytime. On the other hand, healthcare networks usually generate less monitoring data over time, and they should not be re-configured until there is sufficient data available, as frequent reconfigurations may result in failures with severe impacts on peoples' health.

## 8.2 Challenges facing the use of Big Data Analytics in Network Design

### 8.2.1 Network size vs Big Data Analytics gains

Depending on the network size, the ease of redesigning a network through the feedback cycle that we mentioned in **Fig. 4** is highly affected by the number of nodes.

For instance, large data streams can be generated from the mass deployment of small Wireless Sensor Networks (WSNs) nodes and IoT [118]. The collected data may not carry a meaningful value until it is effectively analyzed. However, analyzing or mining that immense amount of data demands finely tuned big data analytical capabilities, which turns out to be a challenging task [119]. Furthermore, these massive amounts of data require hierarchal communication and data processing solutions. The planning of such deployments in conjunction with the data processing framework is a challenging task [118].

Comparing optical to IoT networks, the former has a small number of nodes, hence they are easier to redesign, while the latter has a larger number of connected objects, and that can impose a problem.

### 8.2.2 Security and privacy

Users' common patterns can be of great help. Network users can share certain patterns, like downloading some popular videos, retweeting about some certain upcoming game that would take place downtown, or even checking the same online channels. This information can be of a great value when used for network planning or optimization. However, to use this information, access to user data has to be obtained, which is a thought that may cause unrest for many.

When dealing with user data, there is always a flag raised, and that flag carries two issues: these issues are the security and the privacy of the data. This is why big data has to be protected from unauthorized access and release [35].

Big data security is a vital topic. If we want to label a system as "secured", it must meet the data security requirements, which are [120] :

1- *Confidentiality*: This implies the means to protect the data from unapproved disclosure.

2- *Integrity*: This implies the measures taken to protect the data from being modified improperly or without permission.

3- *Availability*: This is the system's ability to prevent and recover from hardware as well as software failures that might result in the database system being unavailable.

Privacy of data is an increasing concern. As a matter of fact, having accessible data does not mean it is ethical to access it [121]. Electronic health records have strict laws that precisely identify what can and cannot be accessed.

As an example, a user's location information can be tracked through cell towers and after a while, "a trail of crumbs" is going to be left by the user that could be used to link the user to their residence or office location, and to eventually determine the user's identity, private health information (e.g., attending a cancer treatment center) or religious preferences (e.g., attending a church) may be discovered by tracking the user's movement over time [122], especially when we take into consideration the close correlation between an individual's identity and their movement patterns [123]. Some user data can be very valuable, for example, the estimated value of all global personal location data could reach $100 billion in revenue during the next 10 years for service providers, and when it comes to consumers and business end users, that figure can reach up to $700 billion [39].

With no obvious and secure way to handle the collected user data, big data analytics cannot be considered a reliable system. The security issues related to big data analytics can be divided into four concerns, starting with an input (e.g., handheld device, sensor, or even IoT device) where protecting the sensors from being compromised by attacks is regarded as an important security issue, as well as the other areas of data analysis, output, and communication with other systems [124]. It should be noted that these concerns are present in all steps throughout the design cycle shown in **Fig 4**.

A solution that has been designed to address the big data security and privacy challenge is the integrated Rule-Oriented Data (iRODS [125]). This novel technology was designed to ensure security and privacy in big data, and it has some technological features such as federated data grid or "intelligent clouds", distributed rule engine, "iCAT" metadata catalogue, storage access layer that facilitates common access, two ways of interfacing graphical and command line, and APIs to interact with the iRODS data grid [35, 125].

In a position paper, the authors of [126] noted a number of privacy-preserving challenges in the realm of big data analytics, and these challenges are classified as follows:

1- Individuals' Interaction:

    a. *Transparency*: Big data analytics is mostly associated with information collection and processing of specific individuals' data. However, this means that each individual is entitled to know about the data processing operations conducted on his/her data, and the challenging part is in allocating that specific piece of information linked to that person's identity

    b. *Individual's Consent*: According to many privacy laws, an individual is entitled to the right to be asked for his/her *informed consent,* and such consent is a way of ensuring the individual is aware of the type of processing that is conducted. This type of consent, along with the explanation it requires is in fact considered challenging.

    c. *Consent Cancellation and Discarding Personal Data*: Granting consent, on one hand, should also allow the right of revoking it. However, if an individual wished for his/her consent to be canceled, then this means all personal data has to be erased as well. This is a challenging requirement when considering the fact that the data might have been spread to various data collectors and data analysts.

2- Re-Identification Attacks: A user's identity may be compromised when correlating different types of datasets, and this type of attack was further classified:

    a. *Correlation Attacks*.

    b. *Arbitrary Identification Attacks*.

    c. *Targeted Identification Attacks*.

3- Probable vs. Provable Results: Different results can be produced by different queries conducted upon datasets. In this way, a provable link can turn out to be merely a probable one.

4- Economical Outcomes: Providing huge amounts of datasets in advance is essential for big data analytics to work. One way to provide such datasets is by buying them from data providers who offer to sell their users' data to their customers, thus privacy threats might appear. Context faults along with confusion and distraction are just two examples of other threats (i.e., fraud, censorship, and surveillance).

### 8.2.3 Data center scalability

In the big data paradigm, data centers are not only a platform to concentrate data storage, but can also carry out further responsibilities, such as acquiring, managing, organizing, processing and leveraging data values and functions. That would encourage the growth of the infrastructure and related software [36].

The continuous expansion in data volume, coupled with the ever greater demand for faster processing speeds, and the increasing complexity of Relational Database Management System (RDBMS) are considered the main elements to motivate the hunt for expandable (scalable) data centers to handle the data volume and parallel processing requirements; hence, a number of technical challenges have to be taken into consideration when we try to design a scalable data center that can efficiently store, process, and analyze big data, these challenges can be mapped to the middle octagon (big data cluster) shown in **Fig. 4**, and they are:

- Taking into consideration the variety and sheer volume of the disparate data sources, just collecting and integrating data with scalability from scattered locations is a difficult task to accomplish.

- Massive datasets must be mined by big data analytics at different levels and in either a real time or near real time fashion.

- Massive and heterogeneous datasets are to be stored and managed by big data systems while providing the function and performance guarantees needed in terms of fast retrieval, scalability, and privacy protection. . Facebook is a clear example, in that particular matter as it needs to store, access, and analyze over 30 petabytes of user-generated data [39].

Although some might claim that the current problem is not about storage (large volume), but it is about the online processing ability [11], a scalable data center should also incorporate the ability to have a scalable storage system. Non-volatile memory (NVM) technologies are expected to have a promising role in future memory/storage designs [127].

An ideal storage platform has three vital points (constraints) to meet: it should support efficient data access in case of failure (network partitions and node failures), offer its clients a consistent view of the data, and provides high-availability. However, according to Brewer's CAP theorem [128], this ideal system cannot exists, which is due to the fact that it is impossible for the consistency to be guaranteed and for high-availability to be offered in the presence of network partitions. As a result, one of the above constraints has to be relaxed by distributed storage systems [127].

When it comes to securing the required processing speed, Chip Multiprocessors (CMPs) are expected to be the computational plotter for big data analytics [127]. Targeting the emerging trend, Datacenter-on-Chip (DoC) architectures were proposed by the authors of [39], with four usage models that depend on the state of the consolidating applications, if they were cooperating or not. Key scalability challenges were identified and addressed by cache hierarchies and shortage in performance isolation [127, 129].

## 9. Open research directions

1- *Processing minimization*: The first step in the processing of big data is the collection of data and performing pre-processing. Data cleaning is one form of data pre-processing. One particular example where pre-processing might be implemented is using Computational Radio Frequency Identification (CRFID) sensors. In this approach, wireless sensors can be wirelessly powered using technologies like magnetically powered resonance [130], upon proximity to a moving collector object (e.g., a vehicle). This would enable the movement of some of the pre-processing tasks towards the CRFID sensors' side, thus collecting an already cleaned and reduced amount of data that is transferred to the relay before moving it to the data center for final processing. This would allow more efficiency, reduce the analysis time, allow for better storage utilization, and facilitate real-time analytics. As a result, it would lead to faster decision making and an optimally-optimized network.

2- *Facilitating satellite based Internet connectivity in highly populated and poor areas*: Projects like SpaceX are already emerging with more than 4000 satellites and more than 1$ billion combined funding, the project announced by Elon Musk [131], intends to provide high-speed Internet satellites worldwide. By utilizing big data analytics in the field of satellite communication networks, this will focus more power in a selective fashion. The result is less signal-reception requirements, e.g., smaller antenna size and lower Block Up Converter (BUC) power in the above-mentioned areas. Big data analytics can be used to correlate ground data, e.g., geographical info and weather conditions, along with economy-related data to help identify these areas.

3- *Efficient use of idle time*: Big data analytics can be used by operators to help them run their own data and discover patterns that would facilitate service and network optimization. However, analytics may not be a 24/7 job, especially if it is a batch process. Hence, this would leave the equipment and the software in an idle state. An operator may offer the use of his/her equipment to his/her clients from medium and small sized businesses. They could run their data during the idle time, which would offer better energy utilization, provide big data analytics for everyone, and create another source of money where everyone is benefiting. Game theory approaches can be harnessed here to coordinate resource provisioning among several providers.

4- *Analytics reuse*: Cellular networks have high similarity in terms of equipment capabilities, specifications, subscriber requirements, and subscriber geographic distribution. Those operators can benefit from other operator's big data analytics, thus the result of running the data can be applied directly, or after small modifications. For example by omitting the parts associated with different features of the two networks. This would reduce the purchasing cost, minimize the energy consumption, and reduce the optimization time by adopting a proven solution. Another challenge here is to provide a standard APIs between the different operators' equipment so they can access each other's data in an agreed up on manner.

5- *Big data and IoT node placement*: The main cause for the increase in IoT sensors is the desire to collect more data, which –in turn- will result in reaching an enhanced control or comprehension. According to HP, by 2030, IoT sensors will reach one trillion, and this will make IoT data the most significant part of big data [36]. However, gathering data efficiently requires placing the IoT sensors where they can harvest as much data as possible. Many sensors are simply wasted due to placing them in the wrong location (a location that will not be helpful in providing a valuable amount / type of data). Big data can be used to identify these IoT sensors and simply propose better locations, especially when coupled with other information, like weather conditions, social activities, and movement patterns. To reach the optimal IoT network design, big data analytics can correlate several parameters (e.g., traffic patterns, social events, network parameters, and whether conditions) to determine where the best locations are to place the sensors.

6- *Providing test environments for critical applications*: Collecting large amounts of processed data may not be enough to proceed with network reconfiguration. This has to be considered for some critical applications (e.g., health care, military, and aerospace) where human lives could be jeopardized. The design cycle has to comprise an additional test environment in which the proposed design modification has to undergo a certain test cycle before being put to work, although this might postpone the ratification of the newly-proposed design. There will always be a trade-off between accuracy and speed. It is true that waiting for sufficient data to be accumulated would pay off as a better decision-making step, but that rule is not suitable when it comes to critical applications (e.g., medical networks). The design cycle has to undergo a thorough test first. Identifying these applications and providing suitable test environments is a very important task.

7- *Selecting the most efficient energy source for network nodes*: Another aspect that can be added for a greener network is the ability to selectively utilize energy sources based on the correlation of energy source attributes and their ability to serve a particular task. For example, solar energy source can be ideal for outdoor usage during the day when it is sunny, with a backup plan to switch to other sources (i.e., electrical) during special events or bad weather conditions. This can be the case for IoT devices scattered in a business district, where they are mostly utilized during the day, while running idle after the usual office hours.

## 10. Conclusions

There are many areas in which big data analytics can be utilized in the network design process. The concept of gathering network data and correlating them with user trends and service requirements can indeed create an adaptive and user-centric network design.

Throughout our survey, we noticed a lot of focus on the field of wireless communication networks design using big data. Delving deeper reveals that the field of 5G is getting the majority of the researchers' attention due to the new opportunities it has to offer. The optical networking, inter-DC and SDN fields, on the other hand, have yet further research challenges to tackle. We also note that the integration of SDN and big data analytics would facilitate the perfection of the design cycle. The field of network security also has its share where big data analytics is utilized to detect security threats.

Industrial efforts toward optimizing networks based on big data analytics reflect the increasing trend toward employing AI-like approaches, such as pattern recognition and machine learning for network design.

Some of the considered solutions handle big data in a batch manner while others are capable of performing real-time processing. Handling big data in a batch mode can offer more accurate information at the expense of delayed results due to the size of the processed data, while real-time processing offers fast results at the expense of accuracy. Hence, it would be an application-dependent decision whether to choose the former or the latter option.

We predict that the field of network design based on big data analytics will continue to flourish in the near future as more data are collected from the networks and processed to extract useful information regarding network behavior. In the far future, or maybe quite soon, as some claim, employing quantum computing for machine learning purposes could help in dethroning Moor's law and provide more processing space per unit time. This extra space can be harnessed for big data analytics employed in network design.

## 11. Acknowledgment

The authors would like to acknowledge funding from the Engineering and Physical Sciences Research Council (EPSRC), INTERNET (EP/H040536/1) and STAR (EP/K016873/1) projects.

## 12. References

[1] J. Qadir, N. Ahad, E. Mushtaq, M. Bilal, SDNs, Clouds, and Big Data: New Opportunities, 2014 12th International Conference on Frontiers of Information Technology, IEEE, 2014, pp. 28-33.

[2] R. Tudoran, A. Costan, G. Antoniu, OverFlow: Multi-Site Aware Big Data Management for Scientific Workflows on Clouds, IEEE Transactions on Cloud Computing, 4 (2015) 76-89.

[3] S. Gole, A survey of Big Data in social media using data mining techniques, 2015 International Conference on Advanced Computing and Communication Systems, IEEE, 2015, pp. 1-6.

[4] Z. Nyikes, Z. Rajnai, Big Data , As Part of the Critical Infrastructure, (2015) 217-222.

[5] L. Null, J. Lobur, The essentials of computer organization and architecture, Jones & Bartlett Publishers2014.

[6] J. Shemer, P. Neches, The genesis of a database computer, Computer, 17 (1984) 42-56.

[7] V.R. Borkar, M.J. Carey, C. Li, Big data platforms, XRDS: Crossroads, The ACM Magazine for Students, 19 (2012) 44.

[8] S. Ghemawat, H. Gobioff, S.-t. Leung, The Google File System, (2003).

[9] D.J. DeWitt, B. Gerber, G. Graefe, M. Heytens, K. Kumar, G.A. Muralikrishna, A High Performance Dataflow Database Machine, Computer Science Department, University of Wisconsin1986.

[10] S. Fushimi, M. Kitsuregawa, H. Tanaka, An Overview of The System Software of A Parallel Relational Database Machine GRACE, VLDB, 1986, pp. 209-219.

[11] S. Yin, O. Kaynak, Big Data for Modern Industry :, Proceedings of the IEEE, 103 (2015) 143-146.

[12] A.S. Alghamdi, I. Ahmad, T. Hussain, Big Data for C4I Systems : Goals , Applications , Challenges and Tools, International Conference on Innovative Computing Technology (INTECH), 2015, pp. 89-93.

[13] A. McAfee, E. Brynjolfsson, Big Data. The management revolution, Harvard Buiness Review, 90 (2012) 61-68.

[14] V. Moreno-Cano, F. Terroso-Saenz, A.F. Skarmeta-Gomez, Big data for IoT services in smart cities, 2015 IEEE 2nd World Forum on Internet of Things (WF-IoT), IEEE, 2015, pp. 418-423.

[15] K. Sravanthi, T. Subba Redy, Applications of BIG Data in Various Fields, International Journal of Computer Science and Technologies, 6 (2015) 4629-4632.

[16] Y. Lv, Y. Duan, W. Kang, Z. Li, F.-Y. Wang, Traffic Flow Prediction With Big Data: A Deep Learning Approach, IEEE Transactions on Intelligent Transportation Systems, 16 (2014) 865 - 873.

[17] S. Landset, T.M. Khoshgoftaar, A.N. Richter, T. Hasanin, A survey of open source tools for machine learning with big data in the Hadoop ecosystem, Journal of Big Data, 2 (2015) 24.

[18] H. Baek, Sustainable Development Plan for Korea through Expansion of Green IT: Policy Issues for the Effective Utilization of Big Data, Sustainability, 7 (2015) 1308-1328.

[19] S. Kaisler, F. Armour, J.A. Espinosa, W. Money, Big Data: Issues and Challenges Moving Forward, 2013 46th Hawaii International Conference on System Sciences, (2013) 995-1004.

[20] A. Gani, A. Siddiqa, S. Shamshirband, F. Hanum, A survey on indexing techniques for big data: taxonomy and performance evaluation, Knowledge and Information Systems, 46 (2016) 241-284.

[21] Y. Demchenko, P. Grosso, C. De Laat, P. Membrey, Addressing big data issues in Scientific Data Infrastructure, Proceedings of the 2013 International Conference on Collaboration Technologies and Systems, CTS 2013, IEEE, 2013, pp. 48-55.

[22] J. Andreu-Perez, C.C. Poon, R.D. Merrifield, S.T. Wong, G.Z. Yang, Big data for health., IEEE journal of biomedical and health informatics, 19 (2015) 1193-1208.

[23] L. Zhang, A framework to model big data driven complex cyber physical control systems, 2014 20th International Conference on Automation and Computing, IEEE, 2014, pp. 283-288.

[24] P.D.C.d. Almeida, J. Bernardino, Big Data Open Source Platforms, 2015 IEEE International Congress on Big Data, IEEE, 2015, pp. 268-275.

[25] C. Senbalci, S. Altuntas, Z. Bozkus, T. Arsan, Big data platform development with a domain specific language for telecom industries, 2013 High Capacity Optical Networks and Emerging/Enabling Technologies, HONET-CNS 2013, (2013) 116-120.

[26] B. Fan, S. Leng, K. Yang, A dynamic bandwidth allocation algorithm in mobile networks with big data of users and networks, IEEE Network, 30 (2016) 6-10.

[27] C.-L. I, Y. Liu, S. Han, S. Wang, G. Liu, On Big data Analytics for Greener and Softer RAN, IEEE Access, 3 (2015) 3068-3075.

[28] P. Russom, Big data analytics, TDWI Best Practices Report, (2011) 38.

[29] A. Belle, R. Thiagarajan, S.M. Soroushmehr, F. Navidi, D.A. Beard, K. Najarian, Big Data Analytics in Healthcare, Biomed Res Int, 2015 (2015) 370194.

[30] R. Buyya, K. Ramamohanarao, C. Leckie, R.N. Calheiros, A.V. Dastjerdi, S. Versteeg, Big Data Analytics-Enhanced Cloud Computing: Challenges, Architectural Elements, and Future Directions, (2015) 75-84.

[31] P. Gölzer, L. Simon, P. Cato, M. Amberg, Designing Global Manufacturing Networks Using Big Data, Procedia CIRP, 33 (2015) 191-196.

[32] R. Kapdoskar, S. Gaonkar, N. Shelar, A. Surve, P.S. Gavhane, Big Data Analytics, 4 (2015) 518-520.

[33] C. Hu, H. Li, Y. Jiang, Y. Cheng, P. Heegaard, Deep semantics inspection over big network data at wire speed, IEEE Network, 30 (2016) 18-23.

[34] E. Bastug, M. Bennis, E. Zeydan, M.A. Kader, I.A. Karatepe, A.S. Er, M. Debbah, Big data meets telcos: A proactive caching perspective, Journal of Communications and Networks, 17 (2015) 549-557.

[35] B. Matturdi, X. Zhou, S. Li, F. Lin, Big Data security and privacy: A review, China Communications, 11 (2014) 135-145.

[36] M. Chen, S. Mao, Y. Liu, Big data: A survey, Mobile Networks and Applications, 19 (2014) 171-209.

[37] A. Asahara, H. Hayashi, N. Ishimaru, R. Shibasaki, H. Kanasugi, International standard "OGC® moving features" to address "4Vs" on location bigdata, 2015 IEEE International Conference on Big Data (Big Data), IEEE, 2015, pp. 1958-1966.

[38] L. He, P. Yue, Moving towards intelligent giservices, 2015 IEEE International Geoscience and Remote Sensing Symposium (IGARSS), IEEE, 2015, pp. 1373-1376.

[39] H. Hu, Y. Wen, T.-S. Chua, X. Li, Toward Scalable Systems for Big Data Analytics: A Technology Tutorial, IEEE Access, 2 (2014) 652-687.

[40] B. Cyganek, M. Grana, A. Kasprzak, K. Walkowiak, M. Wozniak, Selected aspects of electronic health record analysis from the big data perspective, 2015 IEEE International Conference on Bioinformatics and Biomedicine (BIBM), IEEE, 2015, pp. 1391-1396.

[41] L. Cui, F.R. Yu, Q. Yan, When big data meets software-defined networking: SDN for big data and big data for SDN, IEEE Network, 30 (2016) 58-65.

[42] Y. Demchenko, E. Gruengard, S. Klous, Instructional Model for Building Effective Big Data Curricula for Online and Campus Education, 2014 IEEE

6th International Conference on Cloud Computing Technology and Science, IEEE, 2014, pp. 935-941.

[43] M.A.-u.-d. Khan, M.F. Uddin, N. Gupta, Seven V's of Big Data understanding Big Data to extract value, Proceedings of the 2014 Zone 1 Conference of the American Society for Engineering Education, IEEE, 2014, pp. 1-5.

[44] M.K. Pusala, M.A. Salehi, J.R. Katukuri, Y. Xie, V. Raghavan, Massive Data Analysis: Tasks, Tools, Applications, and Challenges, Big Data Analytics, Springer2016, pp. 11-40.

[45] S. Pyne, B.P. Rao, S.B. Rao, Big Data Analytics: Methods and Applications, Springer2016.

[46] K. Lee, K. Jung, J. Park, D. Kwon, ARLS: A MapReduce-based output analysis tool for large-scale simulations, Advances in Engineering Software, 95 (2016) 28-37.

[47] T. White, Hadoop: The definitive guide, " O'Reilly Media, Inc."2012.

[48] M. Lemoudden, B.E. Ouahidi, Managing cloud-generated logs using big data technologies, 2015 International Conference on Wireless Networks and Mobile Communications (WINCOM), IEEE, 2015, pp. 1-7.

[49] D. Singh, C.K. Reddy, A survey on platforms for big data analytics, Journal of Big Data, 2 (2014) 8.

[50] A.B. Ayed, M.B. Halima, A.M. Alimi, MapReduce Based Text Detection in Big Data Natural Scene Videos, Procedia Computer Science, 53 (2015) 216-223.

[51] N. Zhu, X. Liu, J. Liu, Y. Hua, Towards a cost-efficient MapReduce: mitigating power peaks for Hadoop clusters, Tsinghua Science and Technology, 19 (2014) 24-32.

[52] Big Data in the Enterprise : Network Design Considerations, White Paper, (2011) 1-33.

[53] K. Shvachko, H. Kuang, S. Radia, R. Chansler, The Hadoop Distributed File System, 2010 IEEE 26th Symposium on Mass Storage Systems and Technologies (MSST), IEEE, 2010, pp. 1-10.

[54] M. Zaharia, M. Chowdhury, M.J. Franklin, S. Shenker, I. Stoica, Spark : Cluster Computing with Working Sets, HotCloud'10 Proceedings of the 2nd USENIX conference on Hot topics in cloud computing, (2010) 10.

[55] N. Marz, Storm-distributed and fault-tolerant realtime computation, 2013, URL http://www.storm-project.net.

[56] M. Isard, M. Budiu, Y. Yu, A. Birrell, D. Fetterly, Dryad: distributed data-parallel programs from sequential building blocks, ACM SIGOPS Operating Systems Review, ACM, 2007, pp. 59-72.

[57] V.K. Vavilapalli, S. Seth, B. Saha, C. Curino, O. O'Malley, S. Radia, B. Reed, E. Baldeschwieler, A.C. Murthy, C. Douglas, S. Agarwal, M. Konar, R. Evans, T. Graves, J. Lowe, H. Shah, Apache Hadoop YARN, Proceedings of the 4th annual Symposium on Cloud Computing - SOCC '13, ACM Press, New York, New York, USA, 2013, pp. 1-16.

[58] P. Zikopoulos, C. Eaton, D. DeRoos, Understanding big data, New York et al: McGraw …, (2012) 166.

[59] F. Morales, M. Ruiz, L. Gifre, L.M. Contreras, V. López, L. Velasco, Virtual network topology adaptability based on data analytics for traffic prediction, Journal of Optical Communications and Networking, 9 (2017) A35-A45.

[60] A. Imran, A. Zoha, Challenges in 5G: how to empower SON with big data for enabling 5G, IEEE Network, 28 (2014) 27-33.

[61] H. Daki, A. El Hannani, A. Aqqal, A. Haidine, A. Dahbi, H. Ouahmane, Towards adopting Big Data technologies by mobile networks operators: A Moroccan case study, Cloud Computing Technologies and Applications (CloudTech), 2016 2nd International Conference on, IEEE, 2016, pp. 154-161.

[62] D.S. Terzi, R. Terzi, S. Sagiroglu, Big data analytics for network anomaly detection from netflow data, International Conference on Computer Science and Engineering (UBMK), IEEE, 2017, pp. 592-597.

[63] Ö.F. Çelebi, E. Zeydan, Ö.F. Kurt, Ö. Dedeoglu, Ö. Ieri, B.A. Sungur, A. Akan, S. Ergut, On use of big data for enhancing network coverage analysis, 2013 20th International Conference on Telecommunications, ICT 2013, IEEE, 2013, pp. 1-5.

[64] I.A. Karatepe, E. Zeydan, Anomaly Detection In Cellular Network Data Using Big Data Analytics, European Wireless 2014; 20th European Wireless Conference; Proceedings of, 2014, pp. 1-5.

[65] E.J. Khatib, R. Barco, P. Munoz, I.D. La Bandera, I. Serrano, Self-healing in mobile networks with big data, IEEE Communications Magazine, 54 (2016) 114-120.

[66] E.J. Khatib, R. Barco, I. Serrano, P. Munoz, LTE performance data reduction for knowledge acquisition, Globecom Workshops (GC Wkshps), 2014, IEEE, 2014, pp. 270-274.

[67] I. de la Bandera, R. Barco, P. Munoz, I. Serrano, Cell Outage Detection Based on Handover Statistics, Communications Letters, IEEE, 19 (2015) 1189-1192.

[68] A. Sahni, D. Marwah, R. Chadha, Real time monitoring and analysis of available bandwidth in cellular network-using big data analytics, Computing for Sustainable Global Development (INDIACom), 2015 2nd International Conference on, (2015) 1743-1747.

[69] J. Liu, F. Liu, N. Ansari, Monitoring and analyzing big traffic data of a large-scale cellular network with Hadoop, IEEE Network, 28 (4) (2014) 32-39.

[70] W. Huang, Z. Chen, W. Dong, H. Li, B. Cao, J. Cao, Mobile Internet big data platform in {China} Unicom, Tsinghua Science and Technology, 19 (2014) 95-101.

[71] Wi-Fi direct | Wi-Fi Alliance, URL: http://www.wi-fi.org/discover-wi-fi/wi-fi-direct.

[72] A. Omar, Improving Data Extraction Efficiency of Cache Nodes in Cognitive Radio Networks Using Big Data Analysis, 2015 9th International Conference on Next Generation Mobile Applications, Services and Technologies, IEEE, 2015, pp. 305-310.

[73] K. Zheng, Z. Yang, K. Zhang, P. Chatzimisios, K. Yang, W. Xiang, Big data-driven optimization for mobile networks toward 5G, IEEE Network, 30 (1) (2016) 44-51.

[74] A. Checko, H.L. Christiansen, Y. Yan, L. Scolari, G. Kardaras, M.S. Berger, L. Dittmann, Cloud RAN for mobile networks—a technology overview, Communications Surveys & Tutorials, IEEE, 17 (2015) 405-426.

[75] K.I. Pedersen, Y. Wang, S. Strzyz, F. Frederiksen, Enhanced inter-cell interference coordination in co-channel multi-layer LTE-advanced networks, Wireless Communications, IEEE, 20 (2013) 120-127.

[76] C.-L. Lee, W.-S. Su, K.-A. Tang, W.-I. Chao, Design of handover self-optimization using big data analytics, The 16th Asia-Pacific Network Operations and Management Symposium, IEEE, 2014, pp. 1-5.

[77] P. Kiran, M.G. Jibukumar, C.V. Premkumar, Resource allocation optimization in LTE-A/5G networks using big data analytics, 2016 International Conference on Information Networking (ICOIN), IEEE, 2016, pp. 254-259.

[78] M. Cayrol, H. Farreny, H. Prade, Fuzzy pattern matching, Kybernetes, 11 (1982) 103-116.

[79] A. Imran, A. Zoha, A. Abu-Dayya, Challenges in 5G: How to Empower SON with Big Data for Enabling 5G, Ieee Network, 28 (2014) 27-33.

[80] G. Widmer, M. Kubat, Learning in the presence of concept drift and hidden contexts, Machine learning, 23 (1996) 69-101.

[81] J.A. Hartigan, M.A. Wong, Algorithm AS 136: A K-Means Clustering Algorithm, Journal of the Royal Statistical Society. Series C (Applied Statistics), 28 (1979) 100-108.

[82] D.M. Blei, A.Y. Ng, M.I. Jordan, Latent dirichlet allocation, The Journal of Machine Learning Research, 3 (2003) 993-1022.

[83] Z. Niu, Y. Wu, J. Gong, Z. Yang, Cell zooming for cost-efficient green cellular networks, IEEE Communications Magazine, 48 (2010) 74-79.

[84] N. McKeown, T. Anderson, H. Balakrishnan, G. Parulkar, L. Peterson, J. Rexford, S. Shenker, J. Turner, OpenFlow: enabling innovation in campus networks, ACM SIGCOMM Computer Communication Review, 38 (2008) 69-74.

[85] H. Cui, Y. Zhang, C. Ma, W. Lai, N.C. Beaulieu, S. Sobolevsky, Y. Liu, Design and Realization of Cognitive Routing Resources Using Big Data Analysis in SDN, 2015 IEEE International Congress on Big Data, 2 (2015) 424-429.

[86] M.V. Neves, C.A.F.D. Rose, K. Katrinis, H. Franke, Pythia: Faster Big Data in Motion through Predictive Software-Defined Network Optimization at Runtime, (2014) 82-90.

[87] P. Costa, A. Donnelly, A. Rowstron, G. O'Shea, Camdoop: Exploiting In-network Aggregation for Big Data Applications, Presented as part of the 9th USENIX Symposium on Networked Systems Design and Implementation (NSDI 12), 2012, pp. 29-42.

[88] P. Costa, A. Donnelly, G. O'shea, A. Rowstron, CamCube: a key-based data center, Microsoft Res., Redmond, WA, USA, Technical Report MSR TR-2010-74, (2010).

[89] Y. Yu, M. Isard, D. Fetterly, M. Budiu, Ú. Erlingsson, P.K. Gunda, J. Currey, DryadLINQ: A System for General-Purpose Distributed Data-Parallel Computing Using a High-Level Language, OSDI, 2008, pp. 1-14.

[90] R. Ramaswami, K.N. Sivarajan, Routing and wavelength assignment in all-optical networks, IEEE/ACM Transactions on Networking (TON), 3 (1995) 489-500.

[91] G. Shen, Y. Li, L. Peng, Almost-optimal design for optical networks with hadoop cloud computing: Ten ordinary desktops solve 500-node, 1000-link, and 4000-request RWA problem within three hours, Transparent Optical

Networks (ICTON), 2013 15th International Conference on, IEEE, 2013, pp. 1-4.

[92] R.G. Michael, S.J. David, Computers and intractability: a guide to the theory of NP-completeness, WH Free. Co., San Fr, (1979).

[93] Y. Li, G. Shen, B. Chen, M. Gao, X. Fu, Applying Hadoop Cloud Computing Technique to Optimal Design of Optical Networks, Asia Communications and Photonics Conference, Optical Society of America, 2015, pp. ASu3H. 3.

[94] G. Shen, Y. Lui, S.K. Bose, "Follow the Sun, Follow the Wind" Lightpath Virtual Topology Reconfiguration in IP Over WDM Network, Journal of Lightwave Technology, 32 (2014) 2094-2105.

[95] C. Wang, G. Shen, S.K. Bose, Distance adaptive dynamic routing and spectrum allocation in elastic optical networks with shared backup path protection, Journal of Lightwave Technology, 33 (2015) 2955-2964.

[96] Y. Li, H. Dai, G. Shen, S.K. Bose, Adaptive FEC selection for lightpaths in elastic optical networks, Optical Fiber Communication Conference, Optical Society of America, 2014, pp. W3A. 7.

[97] A. Aguado, M. Davis, S. Peng, M.V. Alvarez, V. López, T. Szyrkowiec, A. Autenrieth, R. Vilalta, A. Mayoral, R. Muñoz, Dynamic virtual network reconfiguration over SDN orchestrated multitechnology optical transport domains, Journal of Lightwave Technology, 34 (2016) 1933-1938.

[98] B. Stone-Gross, M. Cova, L. Cavallaro, B. Gilbert, M. Szydlowski, R. Kemmerer, C. Kruegel, G. Vigna, Your botnet is my botnet: analysis of a botnet takeover, Proceedings of the 16th ACM conference on Computer and communications security, ACM, 2009, pp. 635-647.

[99] WordPress Sites Targeted by Mass Brute-force Botnet Attack | US-CERT, U.S. Department of Homeland Security Seal. United States Computer Emergency Readiness Team US-CERT, 2013.

[100] K. Singh, S.C. Guntuku, A. Thakur, C. Hota, Big Data Analytics framework for Peer-to-Peer Botnet detection using Random Forests, Information Sciences, 278 (2014) 488-497.

[101] C. Sanders, J. Smith, Applied network security monitoring: collection, detection, and analysis, Elsevier2013.

[102] Y. Liu, S. Guo, S. Hu, T. Rabl, H.-A. Jacobsen, J. Li, J. Wang, Performance Evaluation and Optimization of Multi-dimensional Indexes in Hive, IEEE Transactions on Services Computing, pp (2016) 1-1.

[103] S. Ramírez-Gallego, A. Fernández, S. García, M. Chen, F. Herrera, Big Data: Tutorial and guidelines on information and process fusion for analytics algorithms with MapReduce, Information Fusion, 42 (2018) 51-61.

[104] V.P. Janeja, A. Azari, J.M. Namayanja, B. Heilig, B-dids: Mining anomalies in a Big-distributed Intrusion Detection System, 2014 IEEE International Conference on Big Data (Big Data), IEEE, 2014, pp. 32-34.

[105] V. Ayma, R. Ferreira, P. Happ, D. Oliveira, R. Feitosa, G. Costa, A. Plaza, P. Gamba, Classification Algorithms for Big Data Analysis, a Map Reduce Approach, The International Archives of Photogrammetry, Remote Sensing and Spatial Information Sciences, 40 (2015) 17.

[106] Q. Xu, R. Zheng, W. Saad, Z. Han, Device fingerprinting in wireless networks: Challenges and opportunities, IEEE Communications Surveys & Tutorials, 18 (2016) 94-104.

[107] A.L. Buczak, E. Guven, A survey of data mining and machine learning methods for cyber security intrusion detection, IEEE Communications Surveys & Tutorials, 18 (2016) 1153-1176.

[108] M. Molina, I. Paredes-Oliva, W. Routly, P. Barlet-Ros, Operational experiences with anomaly detection in backbone networks, Computers & Security, 31 (3) (2012) 273-285.

[109] F. Ricciato, Traffic monitoring and analysis for the optimization of a 3G network, IEEE Wireless Communications, 13 (4) (2006) 42-49.

[110] M.S. Parwez, D. Rawat, M. Garuba, Big Data Analytics for User Activity Analysis and User Anomaly Detection in Mobile Wireless Network, IEEE Transactions on Industrial Informatics, 13 (2017) 2058 - 2065.

[111] J. Spiess, Y. T'Joens, R. Dragnea, P. Spencer, L. Philippart, Using big data to improve customer experience and business performance, Bell Labs Technical Journal, 18 (4) (2014) 3-17.

[112] J. Zhong, W. Guo, Z. Wang, Study on network failure prediction based on alarm logs, Big Data and Smart City (ICBDSC), 2016 3rd MEC International Conference on, IEEE, 2016, pp. 1-7.

[113] L.H. Shuan, T.Y. Fei, S.W. King, G. Xiaoning, L.Z. Mein, Network Equipment Failure Prediction with Big Data Analytics, International Journal of Advances in Soft Computing & Its Applications, 8 (3) (2016) 59-69.

[114] K. Yang, R. Liu, Y. Sun, J. Yang, X. Chen, Deep Network Analyzer (DNA): A Big Data Analytics Platform for Cellular Networks, IEEE Internet of Things Journal, 4 (6) (2017) 2019-2027.

[115] Y. Qiao, Z. Lei, J. Yang, G. Cheng, FLAS: Traffic analysis of emerging applications on Mobile Internet using cloud computing tools, Wireless Personal Multimedia Communications (WPMC), 2013 16th International Symposium on, IEEE, 2013, pp. 1-6.

[116] G. Qi, W.-T. Tsai, W. Li, Z. Zhu, Y. Luo, A cloud-based triage log analysis and recovery framework, Simulation Modelling Practice and Theory, 77 (2017) 292-316.

[117] B.H. Park, S. Hukerikar, R. Adamson, C. Engelmann, Big Data Meets HPC Log Analytics: Scalable Approach to Understanding Systems at Extreme Scale, Cluster Computing (CLUSTER), 2017 IEEE International Conference on, IEEE, 2017, pp. 758-765.

[118] C. Jardak, P. Mähönen, J. Riihijärvi, Spatial big data and wireless networks: experiences, applications, and research challenges, IEEE Network, 28 (2014) 26-31.

[119] L. Xu, W. He, S. Li, Internet of Things in Industries: A Survey, IEEE Transactions on Industrial Informatics, PP (2014) 1-11.

[120] E. Bertino, Big Data - Security and Privacy, Proceedings of the 5th ACM Conference on Data and Application Security and Privacy, (2015) 757-761.

[121] K. Crawford, Six provocations for big data, (2011) 1-17.

[122] A. Labrinidis, H.V. Jagadish, Challenges and opportunities with big data, Proceedings of the VLDB Endowment, 5 (2012) 2032-2033.

[123] M.C. González, C.A. Hidalgo, A.-L. Barabási, Understanding individual human mobility patterns, Nature, 453 (2008) 779-782.

[124] C.-W. Tsai, C.-F. Lai, H.-C. Chao, A.V. Vasilakos, Big data analytics: a survey, Journal of Big Data, 2 (2015) 21.

[125] A. Rajasekar, R. Moore, C.-Y. Hou, C.a. Lee, R. Marciano, A. de Torcy, M. Wan, W. Schroeder, S.-Y. Chen, L. Gilbert, P. Tooby, B. Zhu, iRODS Primer: Integrated Rule-Oriented Data System, Synthesis Lectures on Information Concepts, Retrieval, and Services, 2 (2010) 1-143.

[126] M. Jensen, Challenges of Privacy Protection in Big Data Analytics, 2013 IEEE International Congress on Big Data, IEEE, 2013, pp. 235-238.

[127] K. Kambatla, G. Kollias, V. Kumar, A. Grama, Trends in big data analytics, Journal of Parallel and Distributed Computing, 74 (2014) 2561-2573.

[128] E.A. Brewer, Towards robust distributed systems, PODC, 2000.

[129] R. Iyer, Datacenter-on-Chip Architectures Terascale Opportunities and Challenges, Intel Technology Journal, 11 (2007).

[130] T. Imura, Y. Hori, Maximizing Air Gap and Efficiency of Magnetic Resonant Coupling for Wireless Power Transfer Using Equivalent Circuit and Neumann Formula, IEEE Transactions on Industrial Electronics, 58 (2011) 4746-4752.

[131] K. Finley, Internet by Satellite Is a Space Race With No Winners, Wired, 2015.

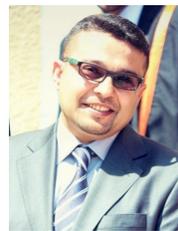

**Mohammed S. Hadi** received the B.Sc. and M.Sc. degrees in computer engineering from Al-Nahrain University, Baghdad, Iraq, in 2003 and 2009 respectively.

He is currently working toward the Ph.D. in Electrical Engineering at the University of Leeds, Leeds, U.K. From (2010 – 2015) he was an assistant lecturer in Al-Mansour University College, Baghdad, Iraq and, prior to that (2007 – 2010), he was an Intelligent Network (IN), Short Message System (SMS), and (Public Switched Telephone Network) PSTN engineer with ZTE Corporation for Telecommunication, Iraq. His research interests include big data analytics, network design and energy efficiency in networks.

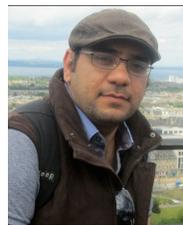

**Ahmed Q. Lawey** received the BS degree (first-class Honors) in computer engineering from the University of Al-Nahrain, Iraq, in 2002, the MSc degree (with distinction) in computer engineering from University of Al-Nahrain, Iraq, in 2005, and the PhD degree in communication networks from the University of Leeds, UK, in 2015.

From 2005 to 2010 he was a core network engineer in ZTE Corporation for Telecommunication, Iraq branch. He is currently a lecturer in communication networks in the School of Electronic and Electrical Engineer, University of Leeds. His current research interests include energy efficiency in optical and wireless networks, big data, cloud computing and Internet of Things.

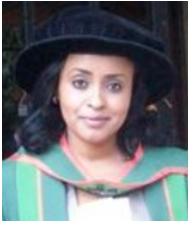

**Taisir E. H. El-Gorashi** received the B.S. degree (first-class Hons.) in electrical and electronic engineering from the University of Khartoum, Khartoum, Sudan, in 2004, the M.Sc. degree (with distinction) in photonic and communication systems from the University of Wales, Swansea, UK, in 2005, and the PhD degree in optical networking from the University of Leeds, Leeds, UK, in 2010. She is currently a Lecturer in optical networks in the School of Electrical and Electronic Engineering, University of Leeds. Previously, she held a Postdoctoral Research post at the University of Leeds (2010–2014), where she focused on the energy efficiency of optical networks investigating the use of renewable energy in core networks, green IP over WDM networks with datacenters, energy efficient physical topology design, energy efficiency of content distribution networks, distributed cloud computing, network virtualization and Big Data. In 2012, she was a BT Research Fellow, where she developed energy efficient hybrid wireless-optical broadband access networks and explored the dynamics of TV viewing behavior and program popularity. The energy efficiency techniques developed during her postdoctoral research contributed 3 out of the 8 carefully chosen core network energy efficiency improvement measures recommended by the GreenTouch consortium for every operator worldwide. Her work led to several invited talks at GreenTouch, Bell Labs, Optical Network Design and Modelling conference, Optical Fiber Communications conference, International Conference on Computer Communications and EU Future Internet Assembly and collaboration with Alcatel Lucent and Huawei.

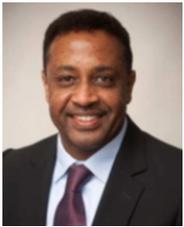

**Jaafar M. H. Elmirghani** (M' 92–SM' 99) is the Director of the Institute of Communication and Power Networks within the School of Electronic and Electrical Engineering, University of Leeds, UK. He joined Leeds in 2007 and prior to that (2000–2007) as chair in optical communications at the University of Wales Swansea he founded, developed and directed the Institute of Advanced Telecommunications and the Technium Digital (TD), a technology incubator/spin-off hub. He has provided outstanding leadership in a number of large research projects at the IAT and TD.

He received the BSc in Electrical Engineering, First Class Honours from the University of Khartoum in 1989 and was awarded all 4 prizes in the department for academic distinction. He received the PhD in the synchronization of optical systems and optical receiver design from the University of Huddersfield UK in 1994 and the DSc in Communication Systems and Networks from University of Leeds, UK, in 2014. He has co-authored Photonic switching Technology: Systems and Networks, (Wiley) and has published over 450 papers. He has research interests in optical systems and networks.

Prof. Elmirghani is Fellow of the IET, Chartered Engineer, Fellow of the Institute of Physics and Senior Member of IEEE. He was Chairman of IEEE Comsoc Transmission Access and Optical Systems technical committee and was Chairman of IEEE Comsoc Signal Processing and Communications Electronics technical committee, and an editor of IEEE Communications Magazine. He was founding Chair of the Advanced Signal Processing for Communication Symposium which started at IEEE GLOBECOM'99 and has continued since at every ICC and GLOBECOM. Prof. Elmirghani was also founding Chair of the first IEEE ICC/GLOBECOM optical symposium at GLOBECOM'00, the Future Photonic Network Technologies, Architectures and Protocols Symposium. He chaired this Symposium, which continues to date under different names. He was the founding chair of the first Green Track at ICC/GLOBECOM at GLOBECOM 2011, and is Chair of the IEEE Green ICT initiative within the IEEE Technical Activities Board (TAB) Future Directions Committee (FDC), a pan IEEE Societies initiative responsible for Green ICT activities across IEEE, 2012-present. He is and has been on the technical program committee of 34 IEEE ICC/GLOBECOM conferences between 1995 and 2016 including 15 times as Symposium Chair. He has given over 55 invited and keynote talks over the past 8 years.

He received the IEEE Communications Society Hal Sobol award, the IEEE Comsoc Chapter Achievement award for excellence in chapter activities (both in international competition in 2005), the University of Wales Swansea Outstanding Research Achievement Award, 2006; and received in international competition: the IEEE Communications Society Signal Processing and Communication Electronics outstanding service award, 2009, a best paper award at IEEE ICC'2013. Related to Green Communications he received (i) the IEEE Comsoc Transmission Access and Optical Systems outstanding Service award 2015 in recognition of "Leadership and Contributions to the Area of Green Communications", (ii) the GreenTouch 1000x award in 2015 for "pioneering research contributions to the field of energy efficiency in telecommunications", (iii) the IET 2016 Premium Award for best paper in IET Optoelectronics and (iv) shared the 2016 Edison Award in the collective disruption category with a team of 6 from GreenTouch for their joint work on the GreenMeter.

He is currently an editor of: IET Optoelectronics and Journal of Optical Communications, and was editor of IEEE Communications Surveys and Tutorials and IEEE Journal on Selected Areas in Communications series on Green Communications and Networking. He was Co-Chair of the GreenTouch Wired, Core and Access Networks Working Group, an adviser to the Commonwealth Scholarship Commission, member of the Royal Society International Joint Projects Panel and member of the Engineering and Physical Sciences Research Council (EPSRC) College. He has been awarded in excess of £22 million in grants to date from EPSRC, the EU and industry and has held prestigious fellowships funded by the Royal Society and by BT. He was an IEEE Comsoc Distinguished Lecturer 2013-2016.